\documentclass[preprint,prd,aps,showpacs,preprintnumbers,amsmath,amssymb]{revtex4-1}
\usepackage{graphics,epsfig,subfigure}
\usepackage{diagbox}
\usepackage{CJKutf8}
\usepackage[usenames]{color}
\usepackage[colorlinks,
            linkcolor=black,
            anchorcolor=blue,
            citecolor=blue
            ]{hyperref}
\usepackage{color}
\usepackage{graphicx}
\usepackage{amsfonts}
\usepackage{indentfirst}
\usepackage{booktabs}

\newcommand\beq{\begin{equation}}
\newcommand\eeq{\end{equation}}
\newcommand\beqn{\begin{eqnarray}}
\newcommand\eeqn{\end{eqnarray}}

\begin{document}
\begin{CJK*}{UTF8}{gbsn}

\title{\Large \bf Bardeen Spacetime with Charged Dirac Field}
\author{Shi-Xian Sun$^{1,2,3}$, Long-Xing Huang$^{1,2,3}$, Zhen-Hua Zhao$^{4,}  $\footnote{haozhh78@sdust.edu.cn, corresponding author}, and Yong-Qiang Wang$^{1,2,3,} $\footnote{yqwang@lzu.edu.cn, corresponding author}}

\affiliation{$^{1}$Institute of Theoretical Physics and Research Center of Gravitation, Lanzhou University, Lanzhou 730000, China\\
	$^{2}$Key Laboratory of Quantum Theory and Applications of MoE, Lanzhou University, Lanzhou 730000, China\\
    $^{3}$Lanzhou Center for Theoretical Physics and Key Laboratory of Theoretical Physics of Gansu Province, Lanzhou University, Lanzhou 730000, China\\
     $^{4}$Department of Applied Physics, Shandong University of Science and Technology, Qingdao  266590, China}

\begin{abstract}
In this article, we investigate soliton solutions in a system involving a charged Dirac field minimally coupled to Einstein gravity and the Bardeen field. We analyze the impact of two key parameters on the properties of the solution family: the magnetic charge $p$ of the Bardeen field and the electric charge $q$ of the Dirac field. We discover that the introduction of the Bardeen field alters the critical charge of the charged Dirac field. In reference \cite{Wang:2023tdz}, solutions named frozen stars are obtained when the magnetic charge is sufficiently large and the frequency approaches zero. In this paper, we define an effective frequency and find that, when the magnetic charge is sufficiently large, a frozen star solution can also be obtained, at which point the effective frequency approaches zero rather than the frequency itself.
\end{abstract}

\maketitle

\section{Introduction}\label{Sec1}

With the Event Horizon Telescope capturing the first photograph of a ``black hole" \cite{EventHorizonTelescope:2019dse, EventHorizonTelescope:2019uob, EventHorizonTelescope:2019jan, EventHorizonTelescope:2019ths, EventHorizonTelescope:2019pgp, EventHorizonTelescope:2019ggy}, people belief in their existence has solidified. Black holes, a profound prediction of Einstein's theory of general relativity, have long been theoretically associated with a singularity. At the singularity, all physical laws break down, leading scientists to hope for a unified theory of gravity and quantum mechanics to address the black hole singularity problem. Unfortunately, such a flawless theory of quantum gravity has yet to be realized.

As demonstrated by Hawking and Penrose, singularities are inevitable if matter adheres to the strong energy condition and certain other prerequisites \cite{Penrose:1964wq, Hawking:1970zqf}. However, this also implies that the introduction of a special type of matter that does not satisfy the strong energy condition might eliminate singularities, potentially leading to the formation of Regular Black Holes (RBHs). The first RBH model was proposed by Bardeen in 1968 \cite{Bardeen:1968}. It was not until 30 years later that the specific type of matter needed to form Bardeen black holes, which do not meet the strong energy condition, was discovered. E. Ayon-Beato and A. Garcia found that magnetic monopoles in the context of nonlinear electrodynamics could serve as the material source for Bardeen's RBH \cite{Ayon-Beato:1998hmi, Ayon-Beato:1999kuh, Ayon-Beato:2000mjt}. Recently, it has been discovered that nonlinear electromagnetic fields can provide various matter sources for RBHs \cite{Bronnikov:2000vy, Dymnikova:2004zc, Ayon-Beato:2004ywd, Berej:2006cc, Lemos:2011dq, Balart:2014jia, Balart:2014cga, Fan:2016rih, Bronnikov:2017sgg, Junior:2023ixh}.

Since we know which types of matter can form RBHs, we can broaden our scope beyond just solutions with the horizon. In fact, magnetic monopoles in nonlinear electromagnetic fields can also yield horizonless solutions. reference \cite{Wang:2023tdz} has identified that introducing a massive complex scalar field into the Bardeen spacetime results in magnetic charge having an upper limit, thus precluding the formation of solutions with the horizon. Moreover, as the frequency of the scalar field approaches zero, a distinct type of solution emerges. In these solutions, there is a special location $ r_{cH} $ within which the scalar field is distributed. For the metric field, $ g_{tt} $ approaches zero for $ r < r_{cH} $, and $ 1/g_{rr} $ is very close to zero at $ r_{cH} $, so we refer to $ r_{cH} $ as the critical horizon. It is noteworthy that because $ 1/g_{rr} $ is only very close to zero, and not exactly zero, this solution does not possess an event horizon and is not classified as a black hole but rather as a ``frozen star". These solutions can also be generalized, such as by replacing the Bardeen background with a Hayward background \cite{Yue:2023sep, Chen:2024bfj}, using different matter fields \cite{Huang:2023fnt, Huang:2024rbg, Zhang:2024ljd}, or considering modified gravity \cite{Ma:2024olw}.

Electromagnetic interactions, being another fundamental force, are often considered in gravitational systems \cite{Reissner:1916cle, Newman:1965my, Hartle:1972ya, Maeda:2008ha, Herdeiro:2018wub}. In classical scenarios, the balance between Newtonian gravity and Coulomb repulsion leads to a critical charge $ q_c $; if a particle’s charge $ q $ exceeds $ q_c $, it cannot condense due to excessive Coulomb repulsion. However, within the framework of general relativity, the gravitational binding energy allows for the existence of soliton solutions with $ q > 1 $. These results have been corroborated by studies on charged boson stars and charged Dirac stars \cite{Jetzer:1989av, Jetzer:1989us, Pugliese:2013gsa, Kumar:2017zms, Collodel:2019ohy, Lopez:2023phk, Jaramillo:2023lgk, Finster:1998ux, Herdeiro:2021jgc}. Research \cite{Huang:2024rbg} has shown that in Bardeen spacetime, the critical charge $ q_c $ for a charged scalar field changes. This paper will explore the interaction between Bardeen spacetime and charged Dirac fields. Similar to previous studies on spherically symmetric Dirac stars \cite{Finster:1998ws, Herdeiro:2017fhv, Dzhunushaliev:2018jhj, Herdeiro:2020jzx, Liang:2023ywv}, this paper considers a pair of Dirac field with opposite spins to maintain spherical symmetry in the spacetime. We find that in Bardeen spacetime, the critical charge for Dirac fields also varies. Moreover, we define an effective frequency and discover that when this effective frequency approaches zero, a frozen star forms.

The structure of this paper is as follows: In section~\ref{Sec2}, we present our model. Section~\ref{Sec3} outlines the boundary conditions that the field functions must satisfy and details our numerical methods. The numerical results are displayed and discussed in section~\ref{Sec4}, followed by a summary in the concluding section.

\section{The Model Setup}\label{Sec2}

We consider a system composed of charged Dirac fields minimally coupled to (3 + 1)-dimensional Einstein's gravity and a nonlinear electromagnetic field (Bardeen field). The action is given by
\begin{equation}
\label{L}
\mathcal{S}=\int d^4 x \sqrt{-g}\left[\frac{R}{16 \pi G}+\mathcal{L}_{B}+\mathcal{L}_{D}+\mathcal{L}_{M}\right],
\end{equation}
where $R$ is the Ricci scalar, $G$ is the gravitational constant. $\mathcal{L}_{B}$, $\mathcal{L}_{D}$ and $\mathcal{L}_{M}$  represent the Lagrangian densities for the Bardeen field, Dirac field, and Maxwell field, respectively, with the following forms:
\begin{equation}
\label{LB}
\mathcal{L}_B = -\frac{3}{2s}\left(\frac{\sqrt{2p^2 \mathcal{H}}}{1+\sqrt{2p^2 \mathcal{H}}}\right)^\frac{5}{2},
\end{equation}
\begin{equation}
\label{LD}
\mathcal{L}_D = -i\sum_{j=1}^{2} \left[\frac{1}{2}(\hat{D}_\mu \bar{\Psi}^{(j)} \gamma^\mu \Psi^{(j)}-\bar{\Psi}^{(j)} \gamma^\mu \hat{D}_\mu \Psi^{(j)})+\mu \bar{\Psi}^{(j)} \Psi^{(j)}\right],
\end{equation}
\begin{equation}
\label{LM}
\mathcal{L}_M =-\frac{1}{4}F_{\mu\nu}F^{\mu\nu}.
\end{equation}
Here, we need to consider two Dirac fields ${\Psi}^{(j)}$ to construct a spherically symmetric spacetime. Two Dirac fields have the same mass $\mu$. $\bar{\Psi}^{(j)}=\Psi^{(j)\dagger}\xi$ are the Dirac conjugate of ${\Psi}^{(j)}$, with $\Psi^{(j)\dagger}$ denoting the usual Hermitian conjugate. For the Hermitian matrix $\xi$, we choose $\xi=-\hat{\gamma}^0$, where $\hat{\gamma}^0$ is one of the gamma matrices in flat spacetime. $\hat{D}_\mu=\partial_\mu-\Gamma_\mu + i q A_\mu$, where $\Gamma_\mu$ are the spinor connection matrices. In equation~(\ref{LB}), $\mathcal{H}\equiv -\frac{1}{4} H_{\mu\nu} H^{\mu\nu}$, the electromagnetic tensors $H_{\mu\nu}$ and $F_{\mu\nu}$ are defined by the electromagnetic four-potentials $B_\mu$ and $A_\mu$, respectively:
\begin{equation}
\label{H}
H_{\mu\nu}\equiv\partial_\mu B_\nu-\partial_\nu B_\mu,
\end{equation}
\begin{equation}
\label{F}
F_{\mu\nu}\equiv\partial_\mu A_\nu-\partial_\nu A_\mu.
\end{equation}

By varying the action, we can obtain the following equations of motion:

\begin{equation}
\label{Eequ}
R_{\mu\nu}-\frac{1}{2} g_{\mu\nu}R-8\pi G T_{\mu\nu}=0,
\end{equation}
\begin{equation}
\label{Hequ}
\nabla_\mu\left(\frac{\partial\mathcal{L}_{B}}{\partial\mathcal{H}}H^{\mu\nu}\right)=0,
\end{equation}
\begin{equation}
\label{Dequ}
\gamma^\mu\hat{D_\mu}\Psi^{(j)}-\mu\Psi^{(j)}=0,
\end{equation}
\begin{equation}
\label{Fequ}
\nabla_\mu F^{\mu\nu}=q (j^\nu_{(1)}+j^\nu_{(2)}), \quad j^\nu_{(j)}=\bar{\Psi}^{(j)} \gamma^\nu \Psi^{(j)},
\end{equation}
where, $T_{\mu\nu}= T_{\mu\nu}^B+T_{\mu\nu}^D+T_{\mu\nu}^M$, is the total energy-momentum tensor. The energy-momentum tensors of the Bardeen field, Dirac fields, and the Maxwell field take the following form:

\begin{equation}
\label{TB}
T_{\mu\nu}^B=-\frac{\partial\mathcal{L}_{B}}{\partial\mathcal{H}}H_{\mu\lambda} {H_\nu}^\lambda +g_{\mu\nu}\mathcal{L}_{B},
\end{equation}
\begin{equation}
\label{TD}
T_{\mu\nu}^D=\sum_{j=1}^{2}\frac{i}{4}(\hat{D}_\mu \bar{\Psi}^{(j)} \gamma_\nu \Psi^{(j)}+\hat{D}_\nu \bar{\Psi}^{(j)} \gamma_\mu \Psi^{(j)}-\bar{\Psi}^{(j)} \gamma_\mu \hat{D}_\nu \Psi^{(j)}-\bar{\Psi}^{(j)} \gamma_\nu \hat{D}_\mu \Psi^{(j)}),
\end{equation}
\begin{equation}
\label{TM}
T_{\mu\nu}^M=-F_{\mu\lambda} {F_\nu}^\lambda -\frac{1}{4}g_{\mu\nu}F_{\lambda\rho}F^{\lambda\rho}.
\end{equation}

In this article, we focus on spherically symmetric spacetime, thus we adopt the following metric ansatz:
\begin{equation}
\label{ds2}
d s^2=-e^{2 F_0(r)} d t^2+e^{2 F_1(r)}d r^2+r^2 d \theta^2+r^2\sin^2\theta d \varphi^2.
\end{equation}
The metric functions $F_0(r)$ and $F_1(r)$ depend only on the radial distance $r$. Additionally, we choose the ansatz for the Dirac fields as follows:
\begin{equation}
\begin{gathered}
\label{p}
\Psi^{(1)}=\left(\begin{array}{c}
z(r) \\
\bar{z}(r)
\end{array}\right) \otimes\left(\begin{array}{c}
-i\cos{\frac{\theta}{2}} \\
\sin{\frac{\theta}{2}}
\end{array}\right) e^{i(\frac{1}{2} \varphi-\omega t)} ,\\
\Psi^{(2)}=\left(\begin{array}{c}
z(r) \\
\bar{z}(r)
\end{array}\right) \otimes\left(\begin{array}{c}
i\sin{\frac{\theta}{2}} \\
\cos{\frac{\theta}{2}}
\end{array}\right) e^{i(-\frac{1}{2} \varphi-\omega t)} .
\end{gathered}
\end{equation}
Here, the constant $\omega$ is the frequency of the Dirac fields, which means that all the spinor fields possess a harmonic time dependence. The part concerning the radial coordinate $r$ is
\begin{equation}
\begin{gathered}
\label{z}
z(r)=ia(r)+b(r),\\
\bar{z}(r)=-ia(r)+b(r).
\end{gathered}
\end{equation}

The real function $a(r)$ and $b(r)$  depend only on the radial distance $r$. For the electromagnetic four-potential $A_\mu$ and $B_\mu$, we employ the following ansatz:
\begin{equation}
\label{A}
A_\mu dx^\mu= c(r) dt,
\end{equation}
\begin{equation}
\label{B}
B_\mu dx^\mu= p \cos{\theta} d\varphi.
\end{equation}
By using the above ansatz, we find that the equations of motion for the Bardeen field naturally hold true. Two Dirac fields satisfy the same equations:
\begin{equation}
\begin{aligned}
\label{ab}
& a^{\prime}+\frac{1}{2} a F_0^{\prime}+\frac{(1+e^{F_1}) a}{r}+e^{F_1} \mu b-e^{F_1-F_0}(\omega-qc)b=0, \\
& b^{\prime}+\frac{1}{2} b F_0^{\prime}+\frac{(1-e^{F_1}) b}{r}+e^{F_1} \mu a+e^{F_1-F_0}(\omega-qc)a=0.
\end{aligned}
\end{equation}
Maxwell's equations read:
\begin{equation}
\label{c}
c^{\prime\prime}+\frac{2c^{\prime}}{r}-(F_0^{\prime}+F_1^{\prime})c^{\prime}+4e^{F_0+2F_1}q(a^2+b^2)=0.
\end{equation}
Equation (\ref{Eequ}) can yield three distinct, independent equations. In our solution process, we utilize two one-order ordinary differential equations concerning the metric functions, as follows:
\begin{equation}
\begin{gathered}
\label{F0F1}
F_0'+\frac{1-e^{2F_1}}{2r}+\frac{6e^{2F_1}Gp^5\pi r}{(p^2+r^2)^{\frac{5}{2}}s}+2e^{-2F_0}G\pi rc'^2+16e^{F_1}G\pi r(ab'-ba')=0,\\
F_1'-\frac{1-e^{2F_1}}{2r}-\frac{6e^{2F_1}Gp^5\pi r}{(p^2+r^2)^{\frac{5}{2}}s}-16e^{2F_1-F_0}G\pi r(\omega-qc)(a^2+b^2)=0,
\end{gathered}
\end{equation}
while the remaining second-order equation serves as a constraint equation to verify the accuracy of the results.

The quantities $j^\mu_{(j)}$ defined in equation~(\ref{Fequ}) are Noether density currents which arise from the invariance of the action~(\ref{L}) under a global $U(1)$ transformation $\Psi^{(j)} \rightarrow \Psi^{(j)}e^{i\alpha}$, where $\alpha$ is constant. This invariance implies the conservation of the total particle number. According to Noether's theorem, we can obtain the particle number:
\begin{equation}
\label{N}
N=\int_{S} J^t_{(1)}+J^t_{(2)},
\end{equation}
here, $S$ is a spacelike hypersurface. Additionally, the total electric charge in spacetime is $Q=qN$. Since the ansatz for the Dirac fields are given by equation~(\ref{p}) and (\ref{z}), it follows that $J^t_{(1)}=J^t_{(2)}$, indicating that the particle numbers for the two Dirac fields are equal. 

ADM mass is also an important global quantity, which can be obtained by integrating the Komar energy density on the spacelike hypersurface $S$:
\begin{equation}
\label{M}
M=\int_{S} T_\mu^\mu-2T_t^t.
\end{equation}

\section{BOUNDARY CONDITIONS AND NUMERICAL METHOD}\label{Sec3}
Before numerically solving these ordinary differential equations, appropriate boundary conditions need to be proposed, which can be determined from the asymptotic behavior of the field functions. Firstly, we require them to be asymptotically flat at spatial infinity ($r\rightarrow \infty$). Thus, we need:
\begin{equation}
\label{inf}
F_0(\infty)=F_1(\infty)=a(\infty)=b(\infty)=0.
\end{equation}
The asymptotic behavior of $c$ cannot be derived from the equations; for convenience, we set the electric potential at spatial infinity to zero, $c(\infty)=0$.
At the origin,  we need:
\begin{equation}
\label{0f}
F_0'(0)=F_1'(0)=a(0)=b'(0)=c'(0)=0.
\end{equation}

There are six input parameters, corresponding to Newton’s constant $G$, the mass of Dirac field $\mu$, the frequency $\omega$, the electric coupling constant $q$, the magnetic charge of the nonlinear electromagnetic field $p$ and the parameter $s$. We take the following field redefinition and scaling of both $r$ and some parameters:
\begin{equation}
\begin{gathered}
\label{sca}
r \rightarrow \frac{r}{\mu},\quad p \rightarrow \frac{p}{\mu},\quad \omega \rightarrow \mu\omega ,\quad q \rightarrow \sqrt{4\pi G} \mu q ,\\
s \rightarrow \frac{4\pi G s}{\mu^2},\quad a \rightarrow \sqrt{\frac{\mu}{4\pi G}}a,\quad b \rightarrow \sqrt{\frac{\mu}{4\pi G}}b,\quad c \rightarrow \sqrt{\frac{1}{4\pi G}}c.
\end{gathered}
\end{equation}
During the solution process, this scaling is equivalent to selecting $\mu = 1$ and $G = \frac{1}{4\pi}$. In all the following results, we set $s=0.3$. As such, we have three input parameters $p$, $q$, and $\omega$. To facilitate this, we introduce a new coordinate $x$, transforming the radial coordinate range from $[0,\infty)$ to $[0, 1]$,
\begin{equation}
\label{rtx}
x=\frac{r}{r+1}.
\end{equation}
Our numerical calculations are all based on the finite element method, with 1000 grid points in the integration domain of $0 \leq x \leq 1$. We employ the Newton-Raphson method as the iterative scheme. To ensure the accuracy of the computed results, we require the relative error to be less than $10^{-5}$.

\section{Numerical results}\label{Sec4}
\subsection{Field Function}\label{sSec1}
\begin{figure}[h!]
    \begin{center}
        \includegraphics[height=.22\textheight]{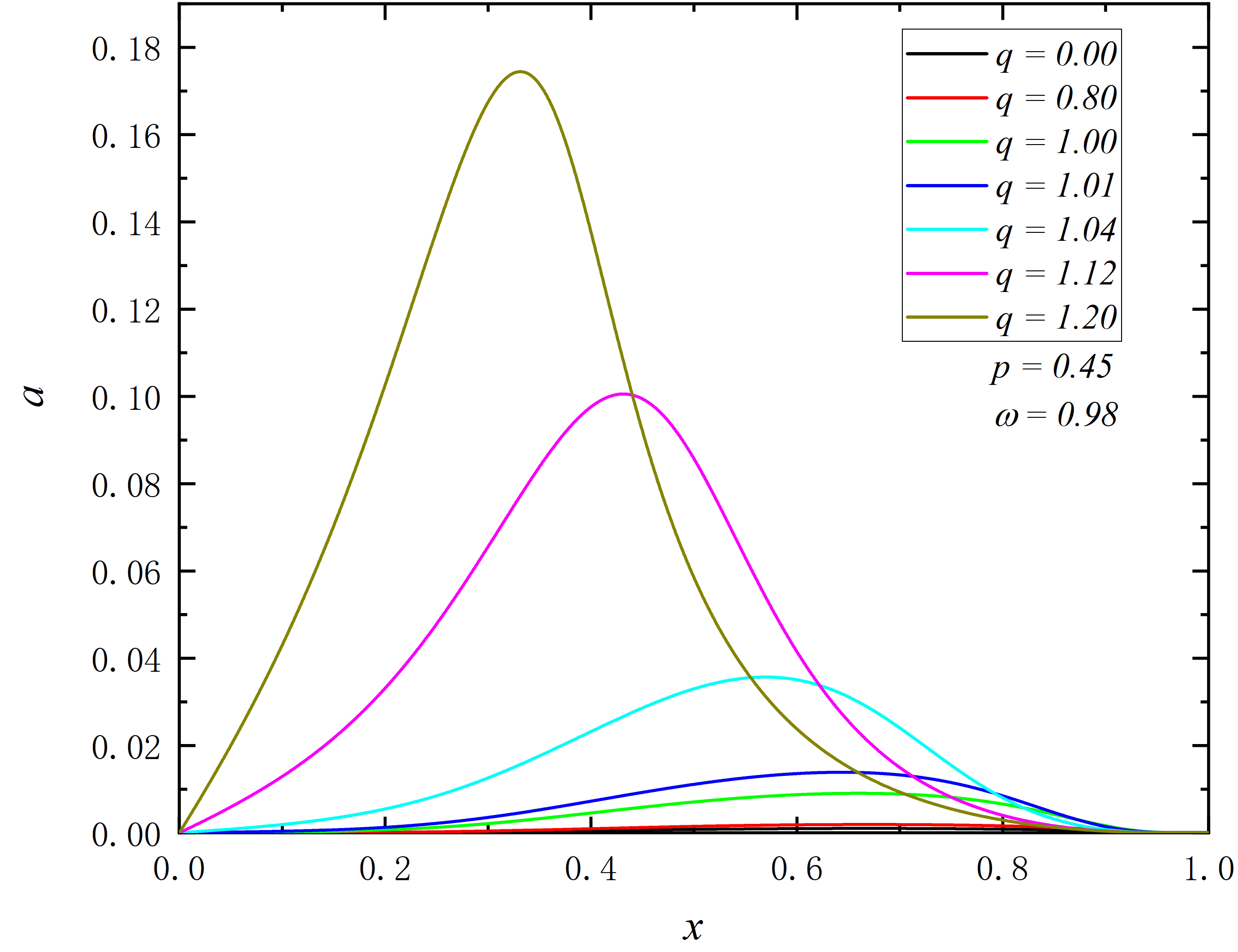}
        \includegraphics[height=.22\textheight]{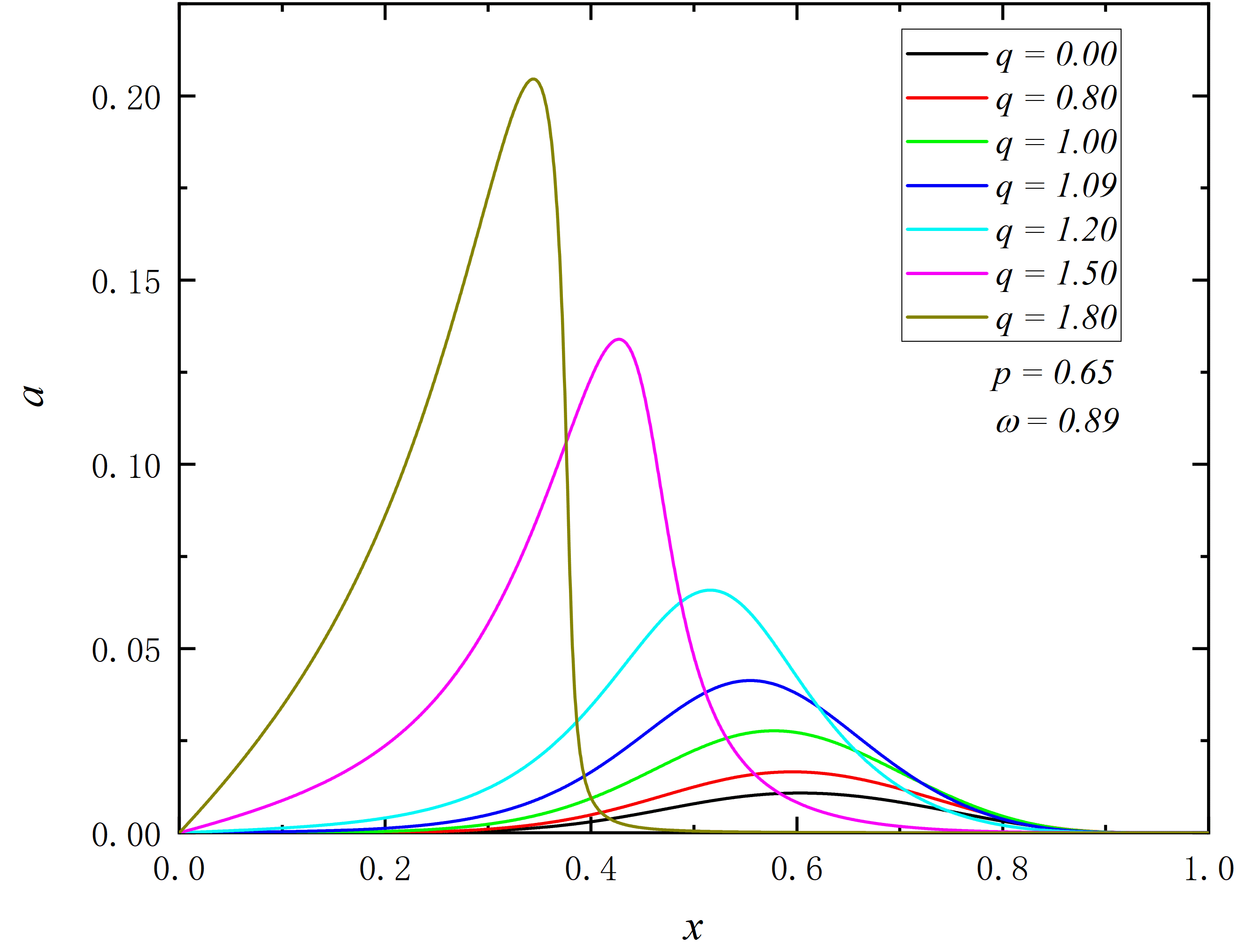}
        \includegraphics[height=.22\textheight]{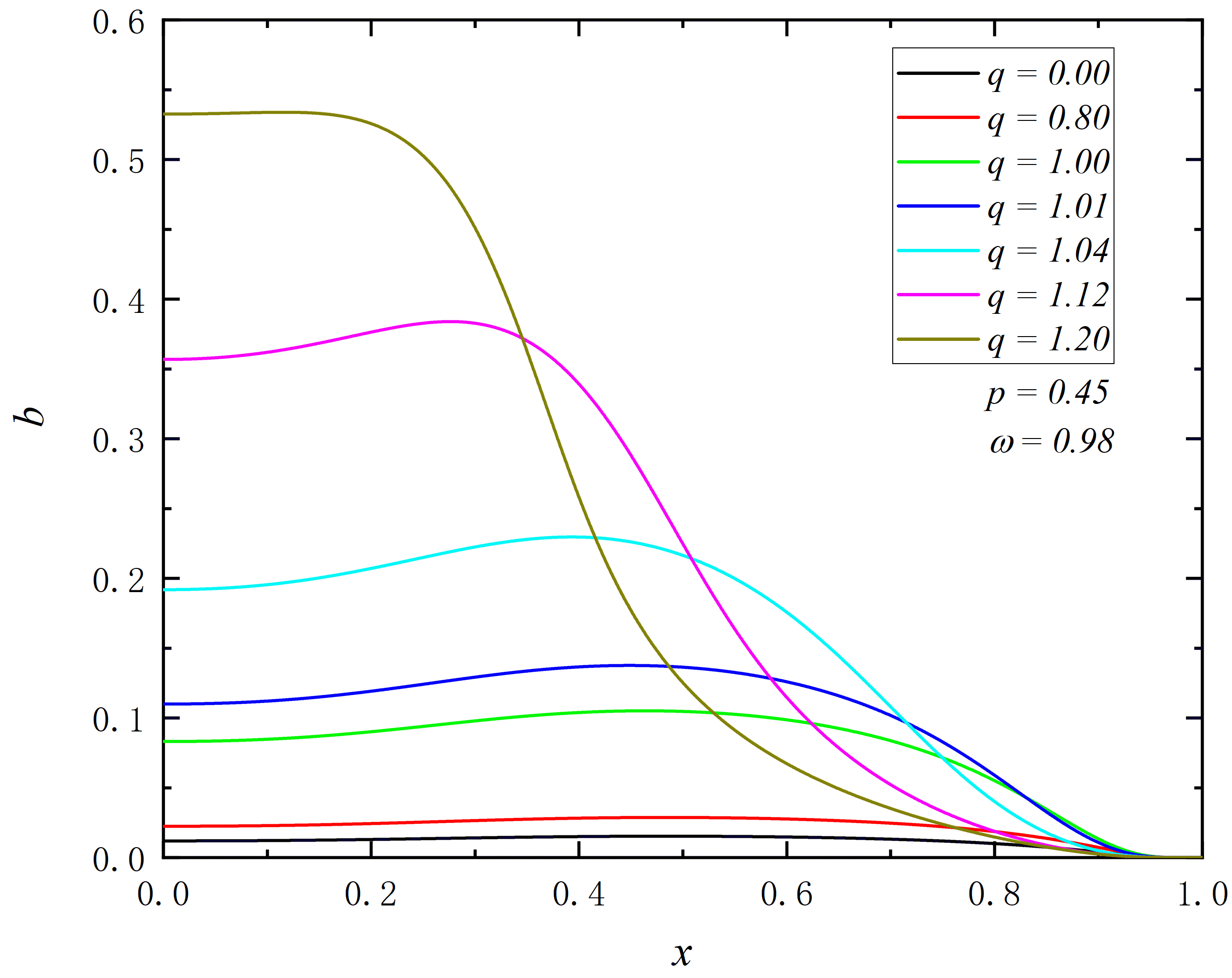}
        \includegraphics[height=.22\textheight]{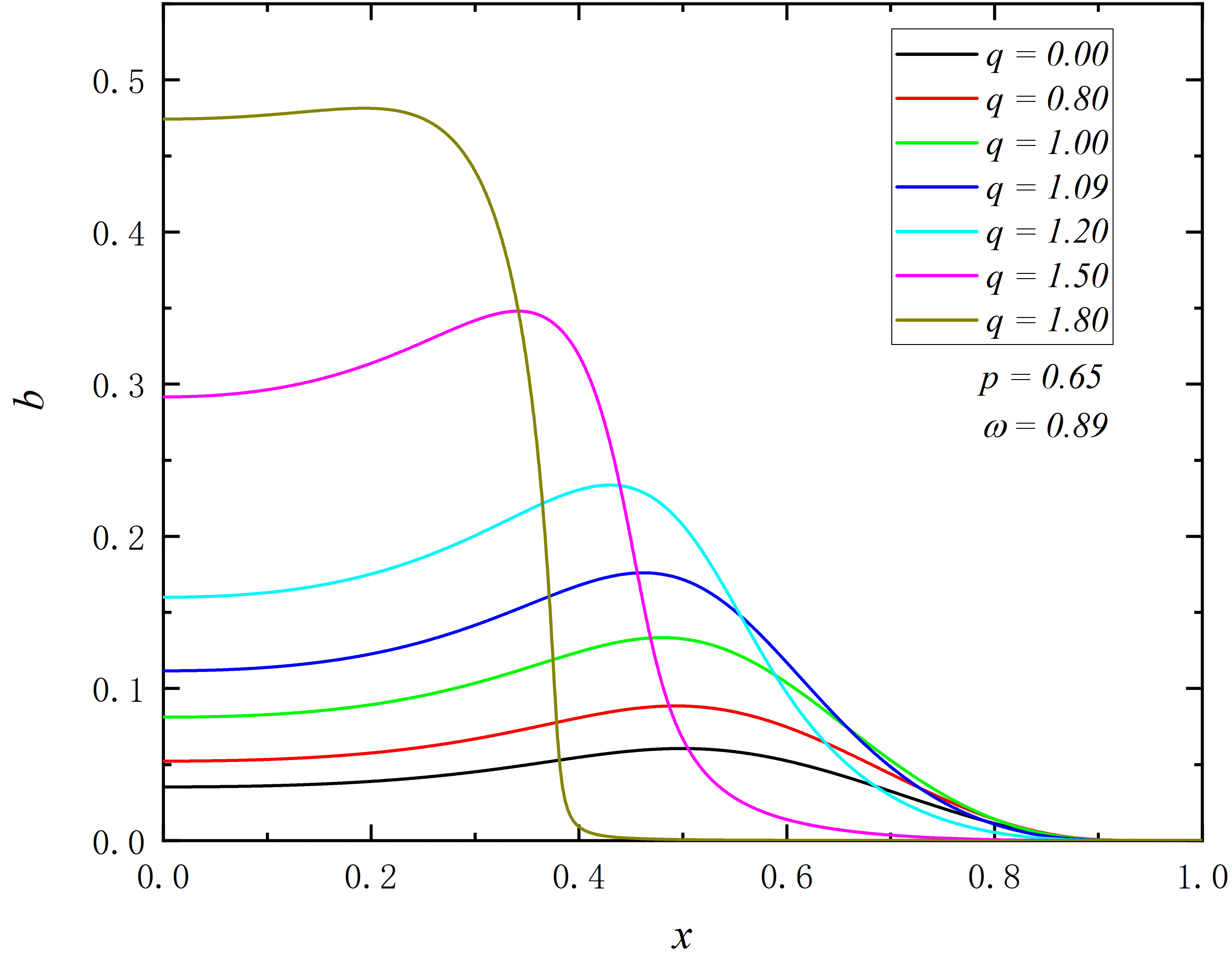}
        \includegraphics[height=.22\textheight]{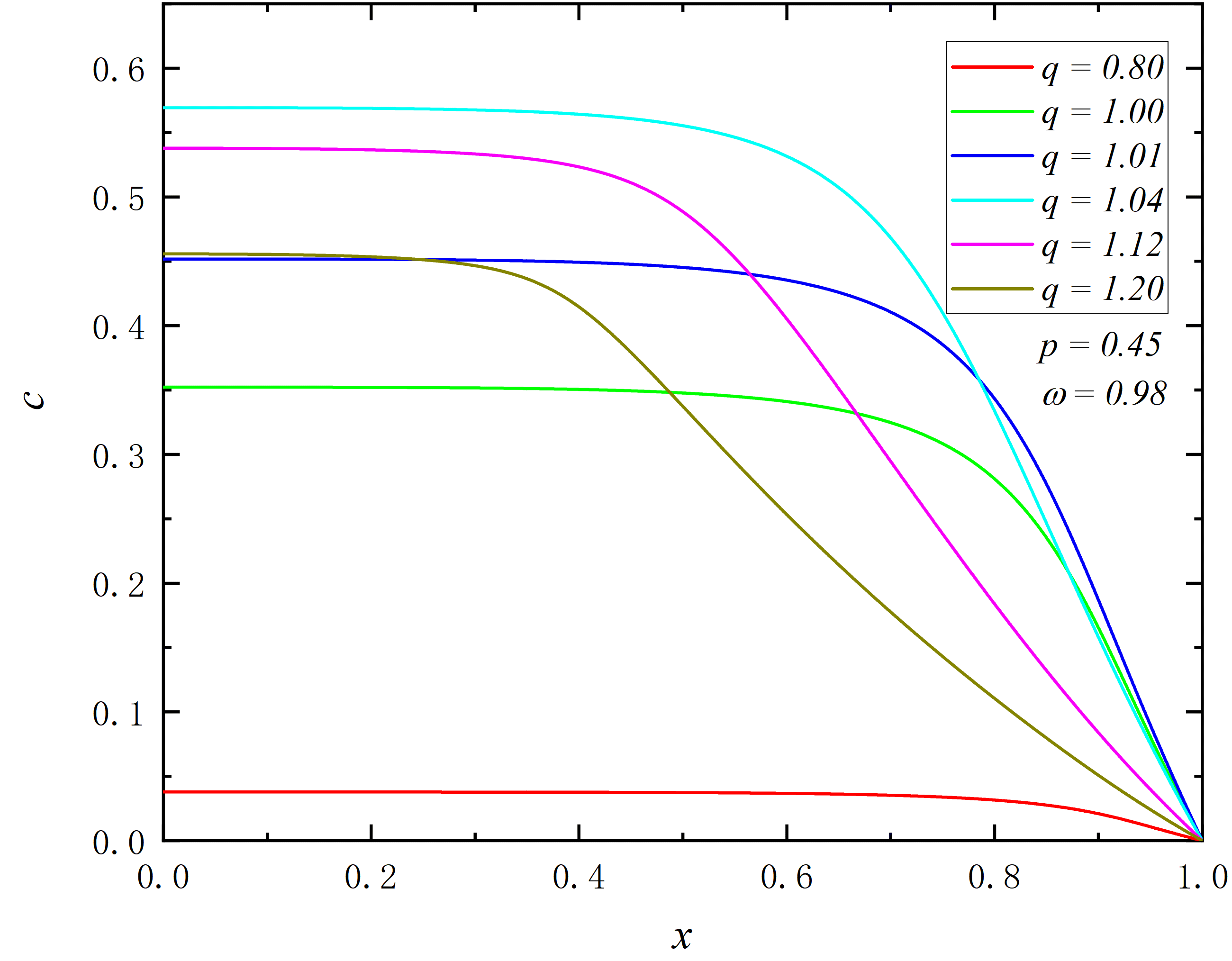}
        \includegraphics[height=.22\textheight]{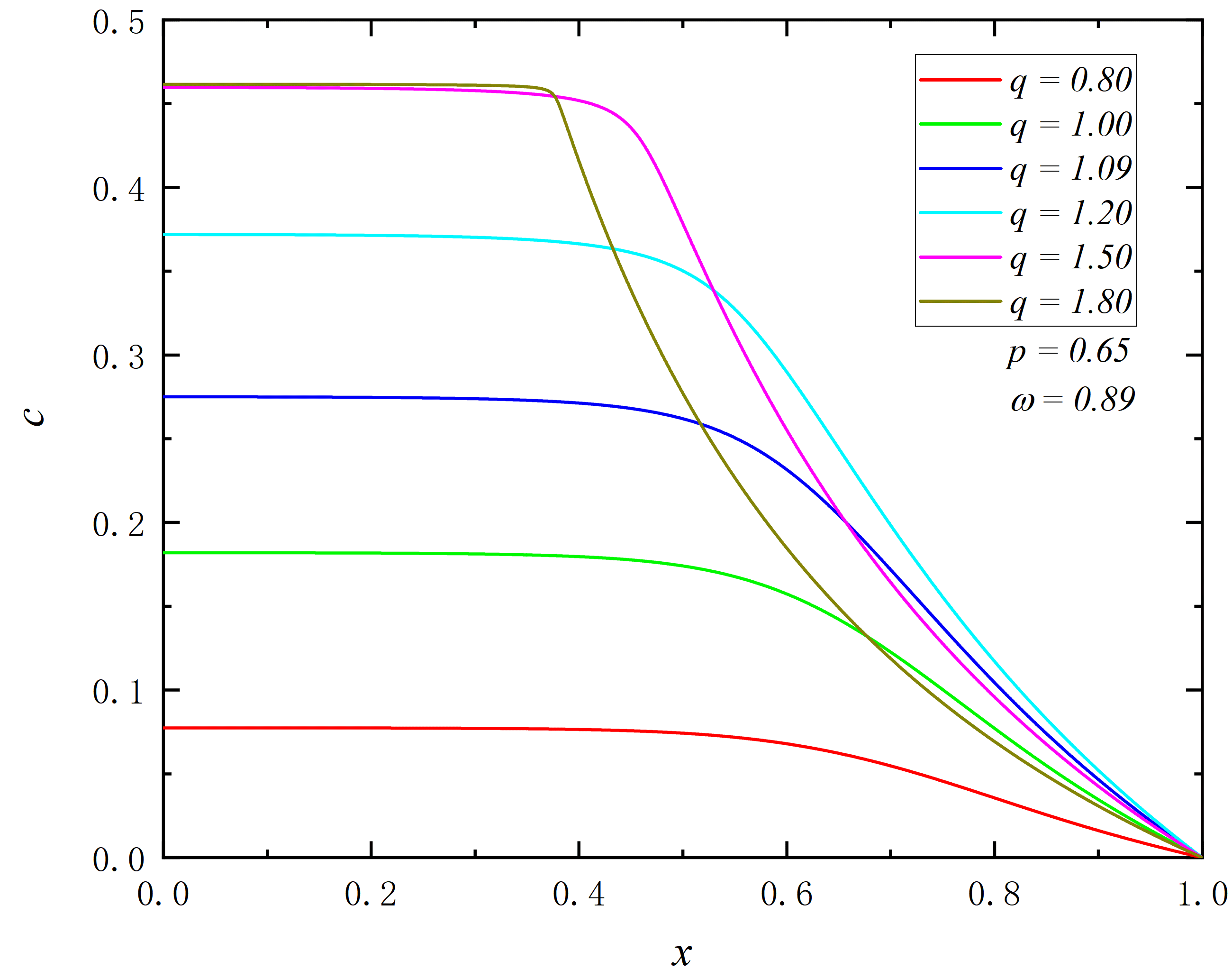}
    \end{center}
    \caption{\small
\textit{Left}: The Dirac field functions $a$, $b$ and the potential function $c$ of the electric field for solutions with $p=0.45$, $\omega=0.98$. \textit{Right}: The Dirac field functions $a$, $b$ and the potential function $c$ of the electric field for solutions with $p=0.65$, $\omega=0.89$. The different colors of the lines represent various values of $q$, as specified in the legend.}
\label{abc}
\end{figure}

To illustrate the impact of the electric coupling constant $q$ on the field functions, we present in figure~\ref{abc} the distributions of the Dirac field functions $a$ and $b$, and the potential function $c$ of the electric field. The left three graphs correspond to $p=0.45$, $\omega=0.98$, while the right three graphs correspond to $p=0.65$, $\omega=0.89$. It is evident that for these solutions, both functions $a$ and $b$ exhibit maximum that are not centered. The maximum value of function $c$ occurs at the center and the value of $c$ decreases gradually from the center to infinity. Additionally, as $q$ increases, the maximum of functions $a$ and $b$ increase, and the location of the maximum moves closer to the center. For function $c$, in the case of solutions with $p=0.65$, $\omega=0.89$, the central value of $c$ increases with $q$. However, for solutions with $p=0.45$, $\omega=0.98$, this is not the case; the central value of $c$ first increases and then decreases with increasing $q$. Upon examining the solutions, we found that for larger values of $p$, the central value of $c$ increases with $q$ for all $\omega$. As $p$ decreases, the frequency range within which the central value of $c$ first increases and then decreases expands.

\subsection{ADM Mass}\label{sSec2}
\begin{figure}[h!]
    \begin{center}
        \includegraphics[height=.22\textheight]{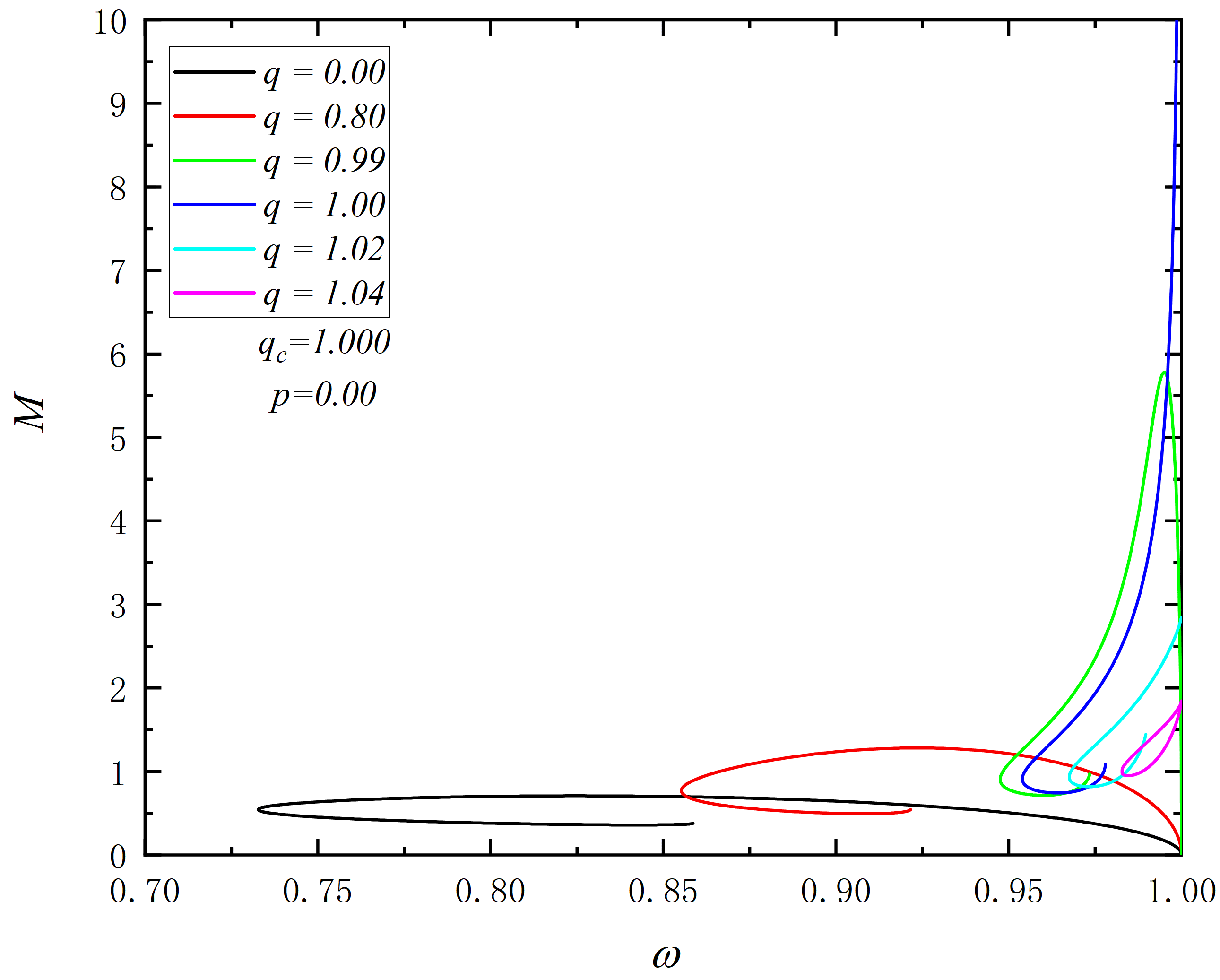}
        \includegraphics[height=.22\textheight]{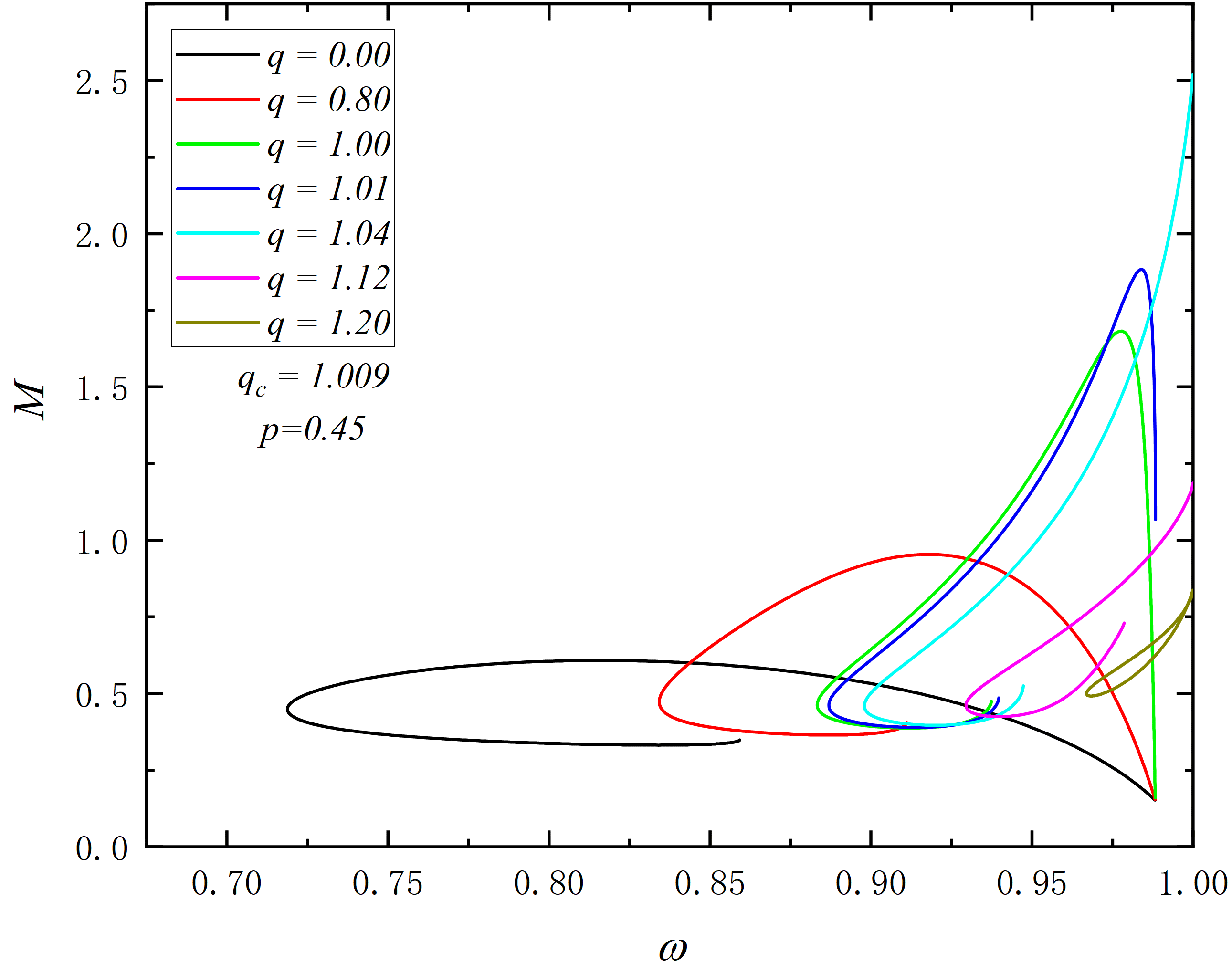}
        \includegraphics[height=.22\textheight]{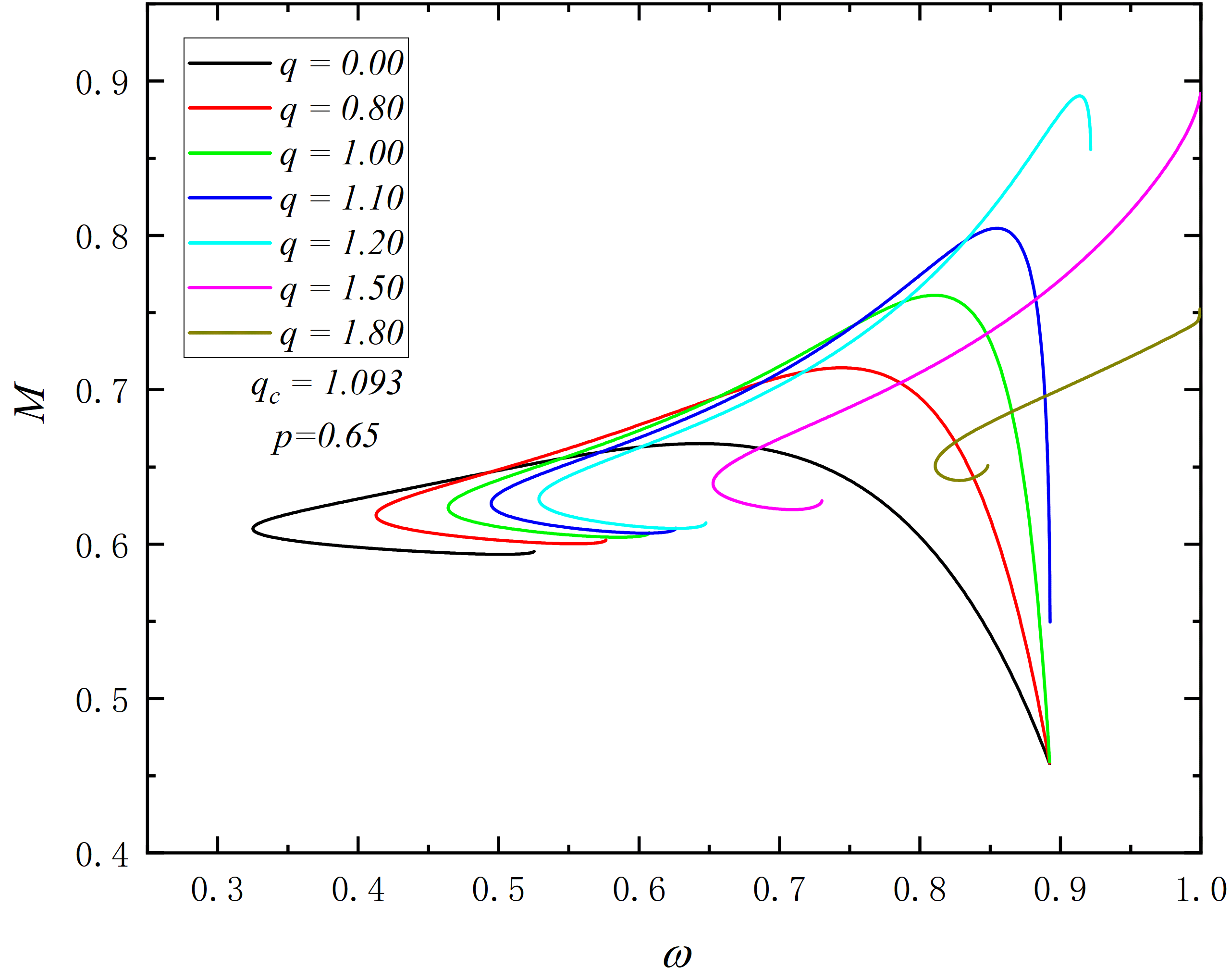}
        \includegraphics[height=.22\textheight]{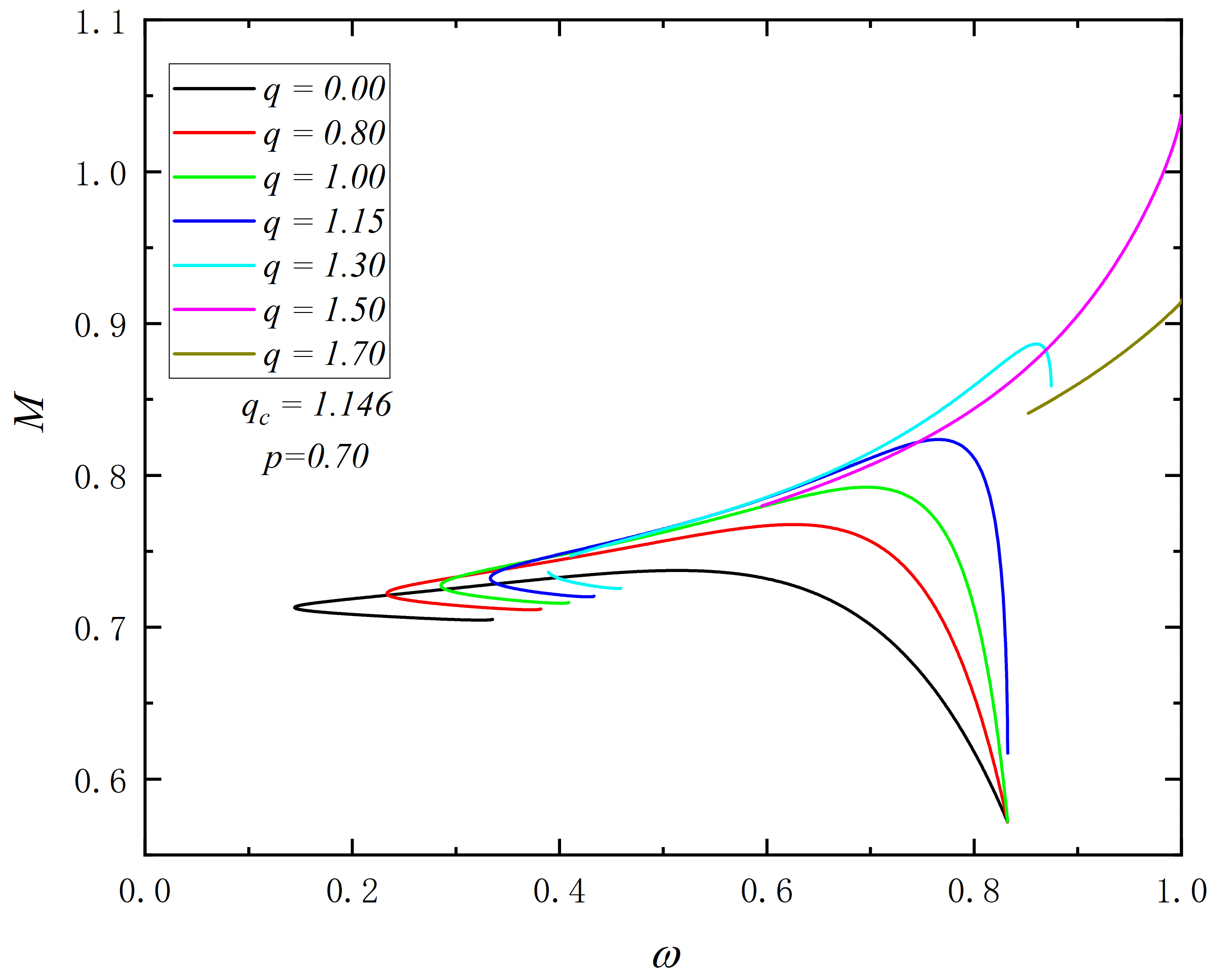}
        \includegraphics[height=.22\textheight]{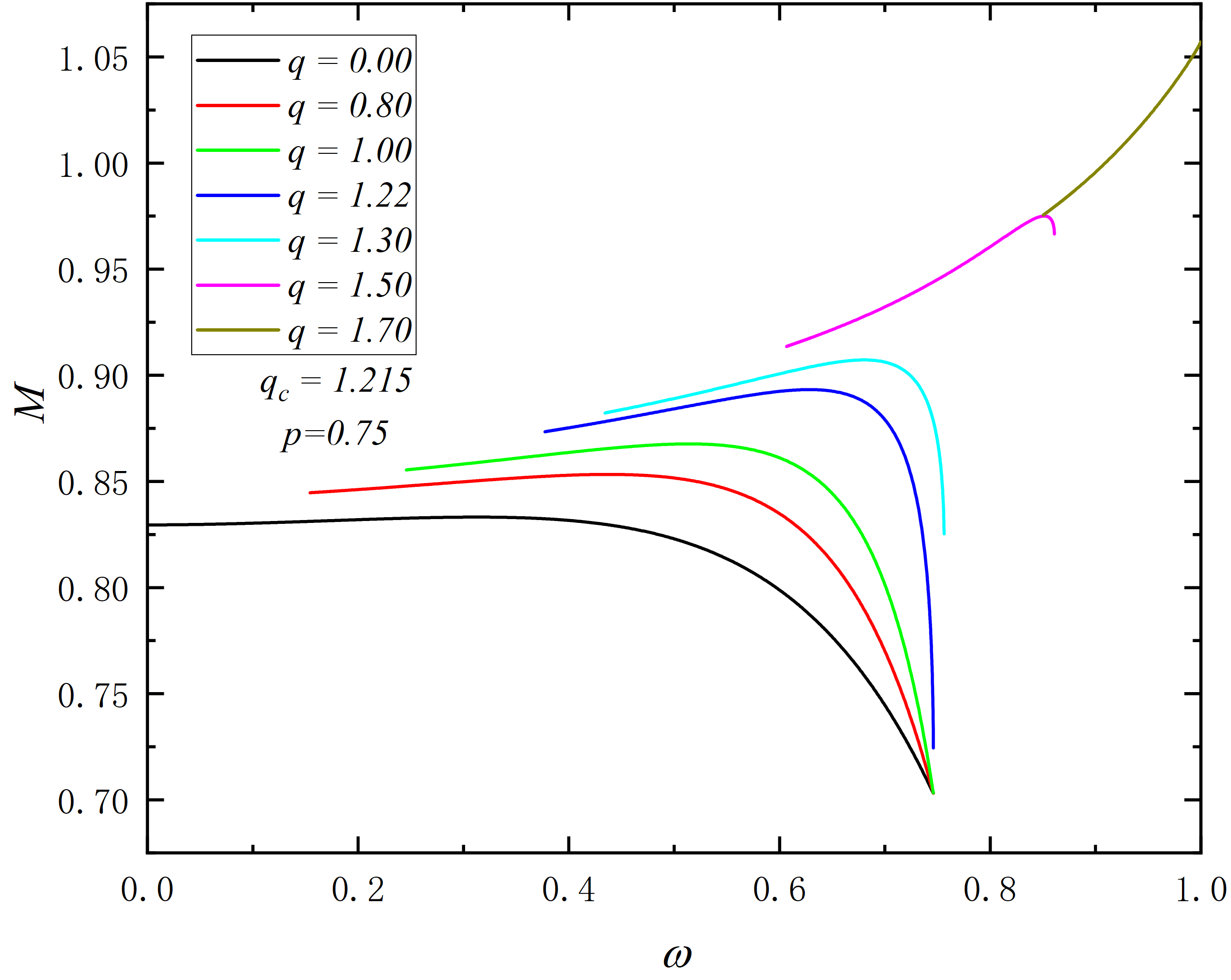}
        \includegraphics[height=.22\textheight]{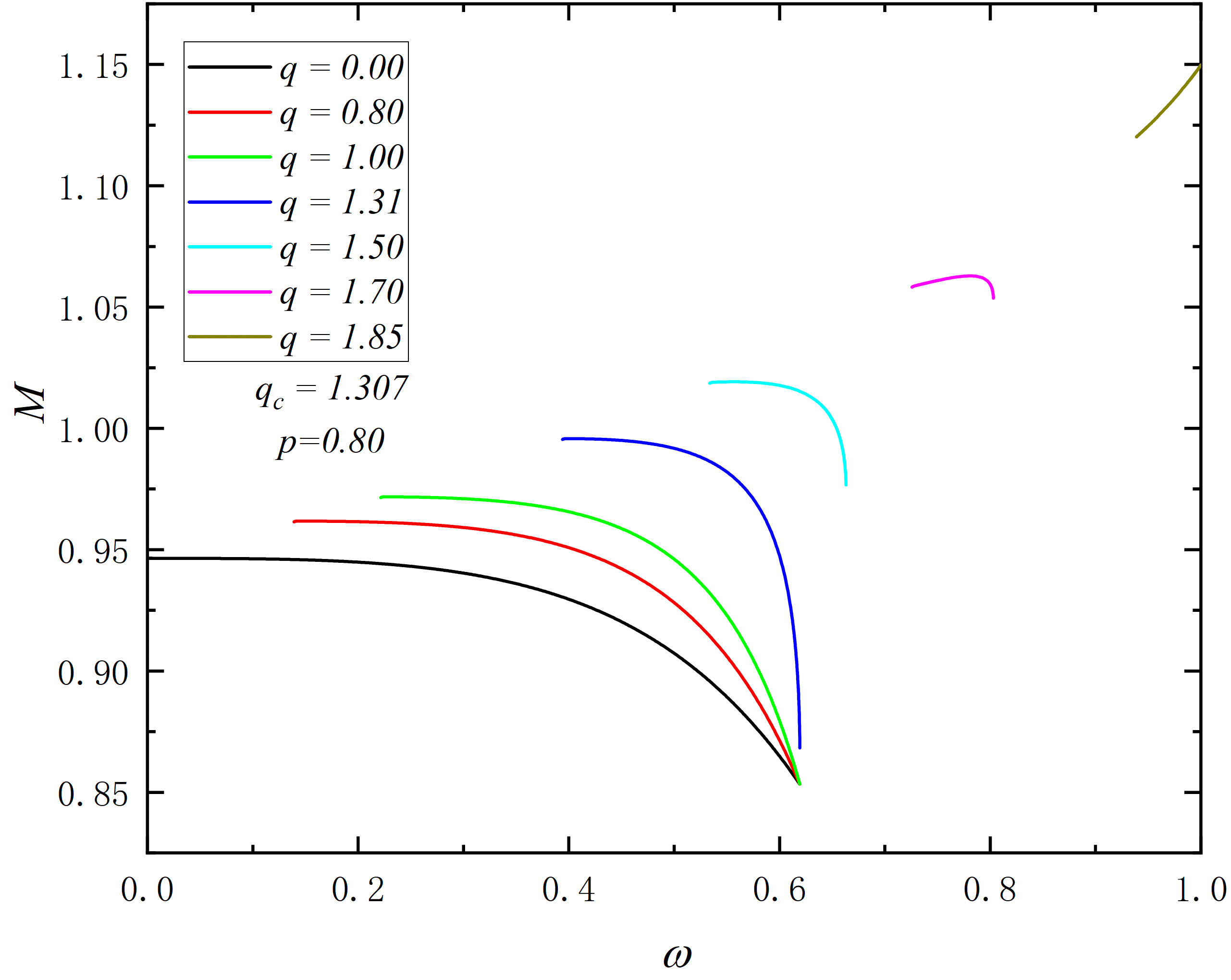}
    \end{center}
    \caption{\small
The ADM mass $M$ as a function of the frequency $\omega$. The different panels correspond to distinct values of $p$, specifically $p=0.00, 0.45, 0.65, 0.70, 0.75, 0.80, 0.85$. The different colors of the lines represent various values of $q$, as specified in the legend. The critical charges $q_c$ corresponding to the different $p$  are also indicated in panels.}
\label{admm}
\end{figure}

In figure~\ref{admm}, we present the relationship between the ADM mass $M$ and frequency for various  $p$ and $q$. In each panel, $p$ is held constant, with different colored curves representing different values of $q$. The curves exhibit a spiraling trend and display different characteristics based on the parameters:

\textbf{i. Critical Charges:} In the upper left panel, where $p = 0$ and $q_c = 1$, for $q < q_c$, as the frequency increases to $1$, the solution returns to the Minkowski vacuum. For $q \geq q_c$, the solution cannot return to the Minkowski vacuum. When $p \neq 0$, we find that $q_c \neq 1$, and the corresponding $q_c$ values are shown in the panel, with precision up to $0.001$. As $p$ increases, the critical charge $q_c$ gradually increases. As the frequency increases to its maximum, if $q < q_c$, the solution reverts to the pure Bardeen spacetime; if $q > q_c$, it cannot return to the pure Bardeen spacetime. Moreover, for any $p$, curves corresponding to $q < q_c$ have overlapping right endpoints (same maximum frequency). When $q > q_c$, the maximum frequency increases with $q$ until $\omega_{max} = 1$.

\textbf{ii. Maximum of Frequency:} From figure~\ref{admm}, for solutions with maximum frequencies less than $1$ and $p \neq 0$, on the first branch of the curve, the ADM mass first increases and then decreases as the frequency decreases, with the maximum ADM mass increasing with $q$. For solutions with a $\omega_{max} = 1$, on the first branch of the curve, the ADM mass monotonically decreases with decreasing frequency, and the maximum ADM mass decreases with increasing $q$.

\textbf{iii. Magnetic Charge:} For smaller values of $p$, the curves exhibit a second branch. As $p$ increases, the minimum frequency gradually decreases, and the second branch eventually disappears. For $q = 0$, as $p$ increases, the minimum frequency decreases to zero, while for $q \neq 0$, the minimum frequency remains non-zero. Additionally, as $p$ increases, $q_c$ also increases, indicating that nonlinear electromagnetic fields provide more gravitational support to counteract the  Coulomb repulsion between charged Dirac fields.

\subsection{Effective Frequency}\label{sSec3}
\begin{figure}[h!]
    \begin{center}
        \includegraphics[height=.22\textheight]{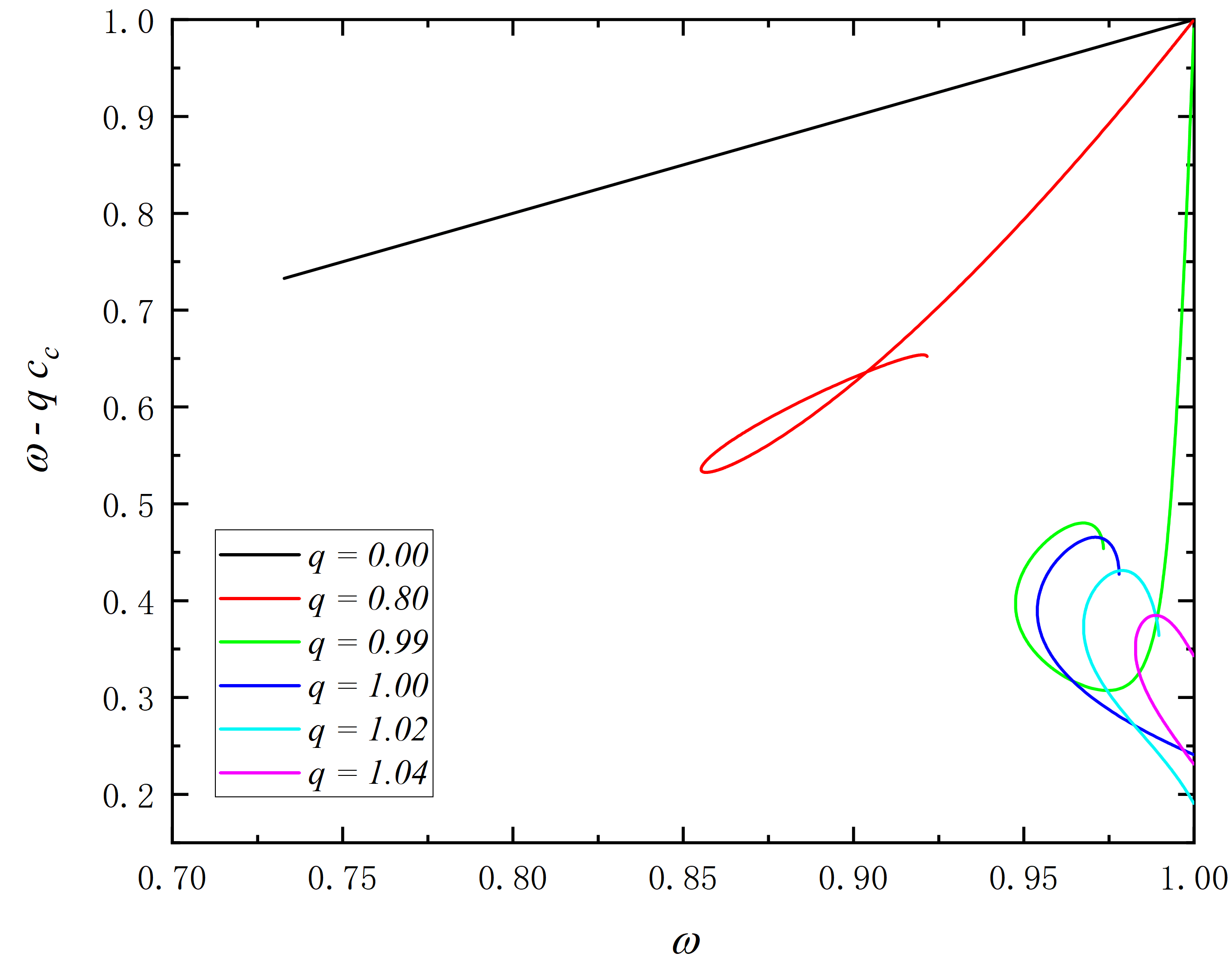}
        \includegraphics[height=.22\textheight]{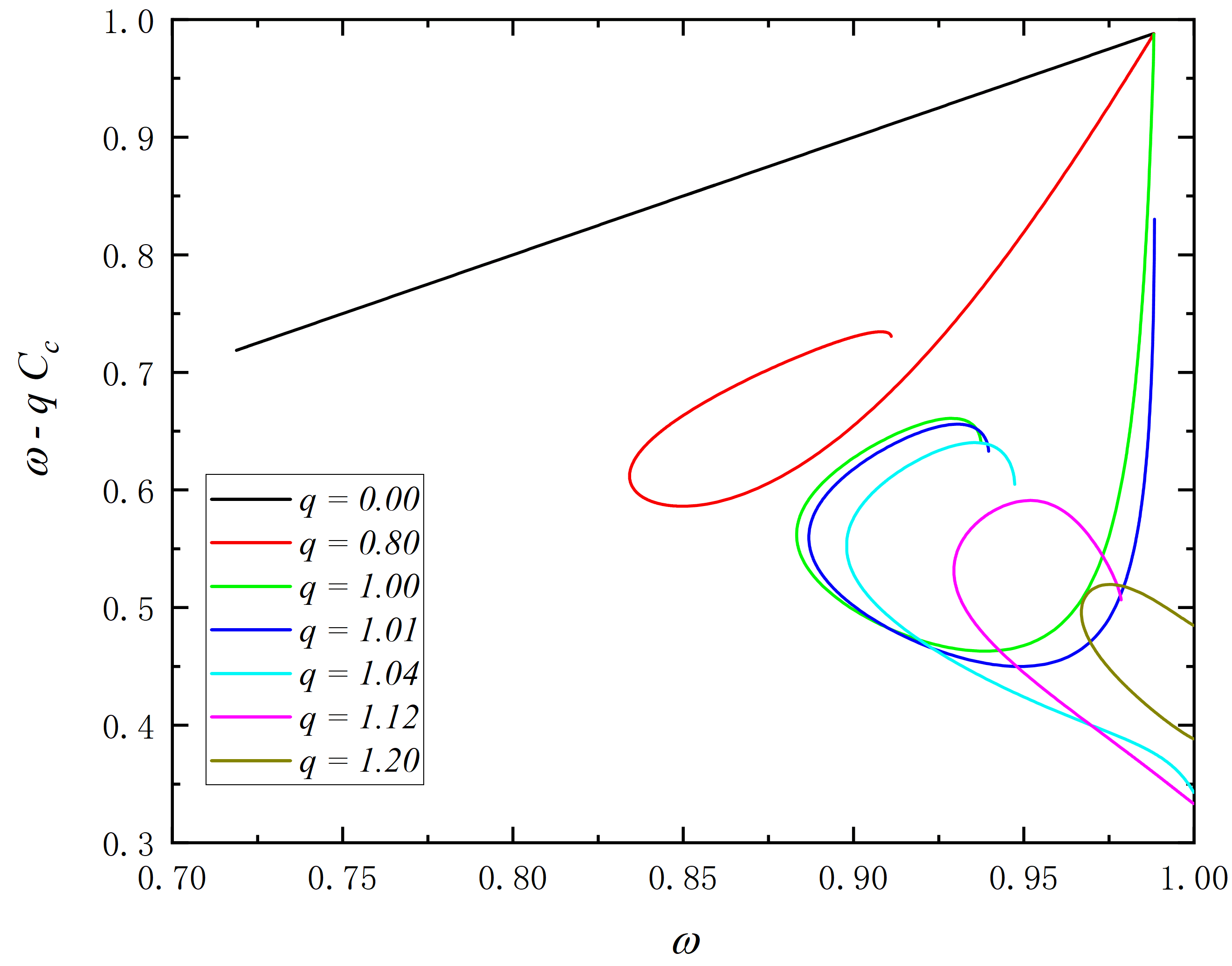}
        \includegraphics[height=.22\textheight]{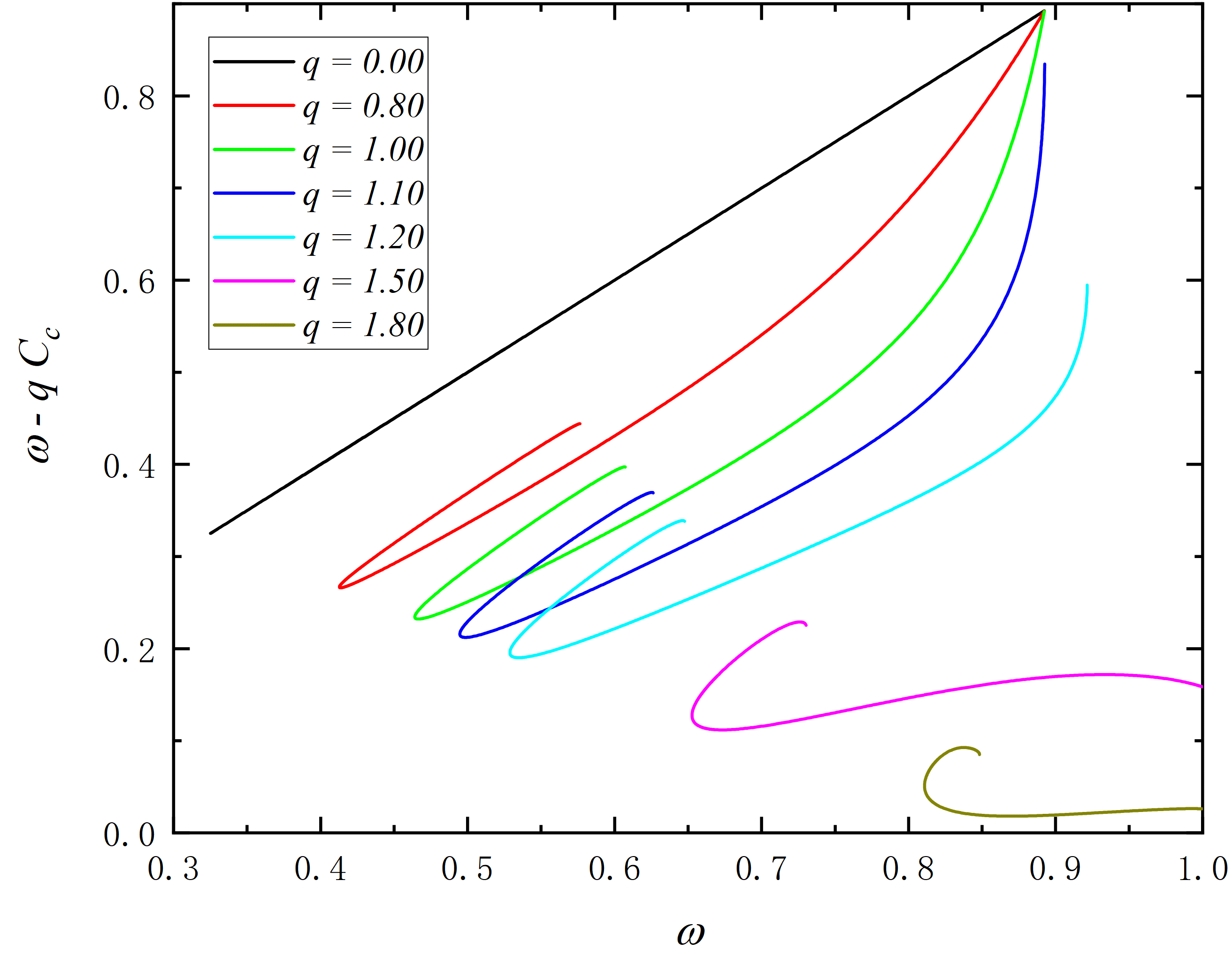}
        \includegraphics[height=.22\textheight]{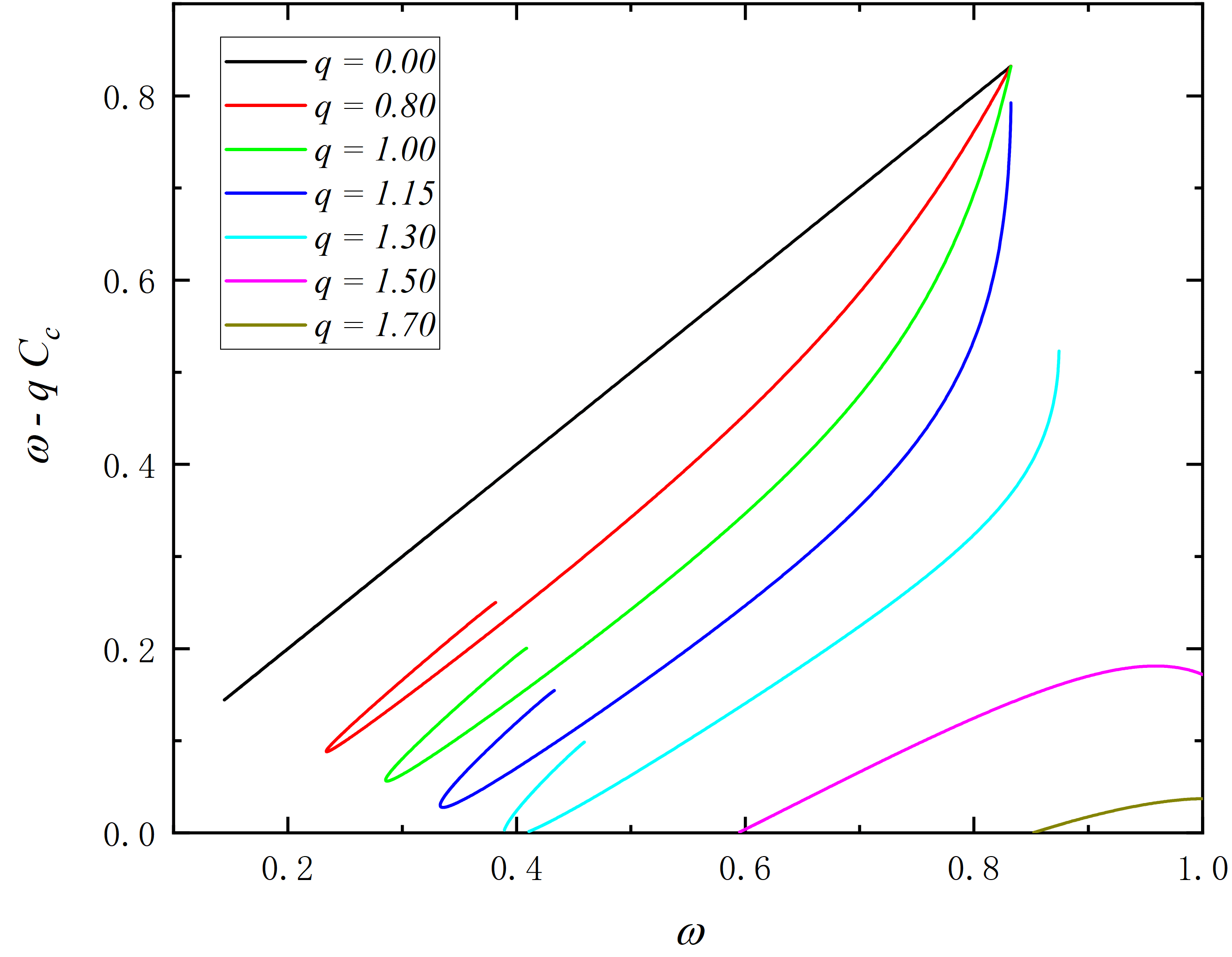}
        \includegraphics[height=.22\textheight]{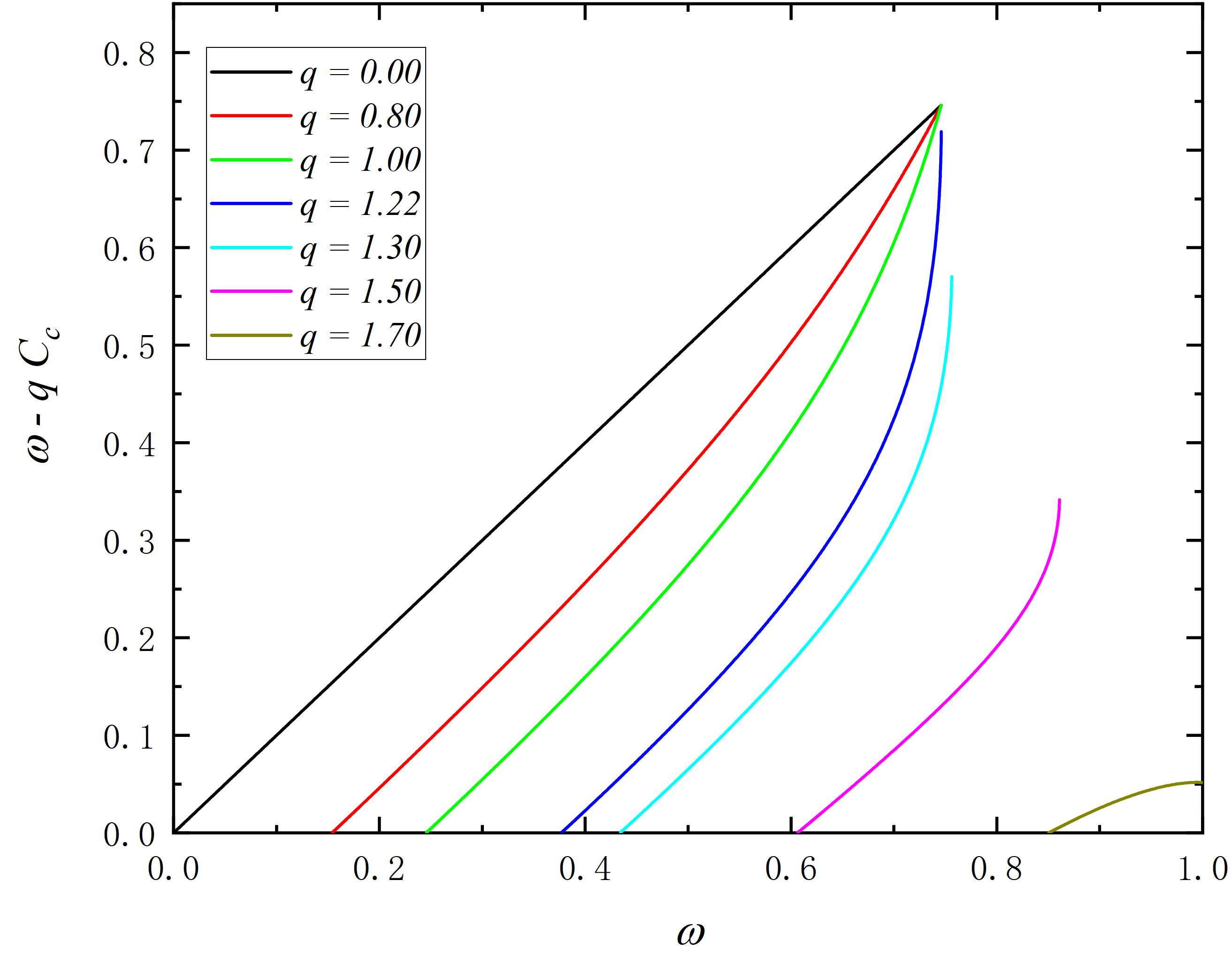}
        \includegraphics[height=.22\textheight]{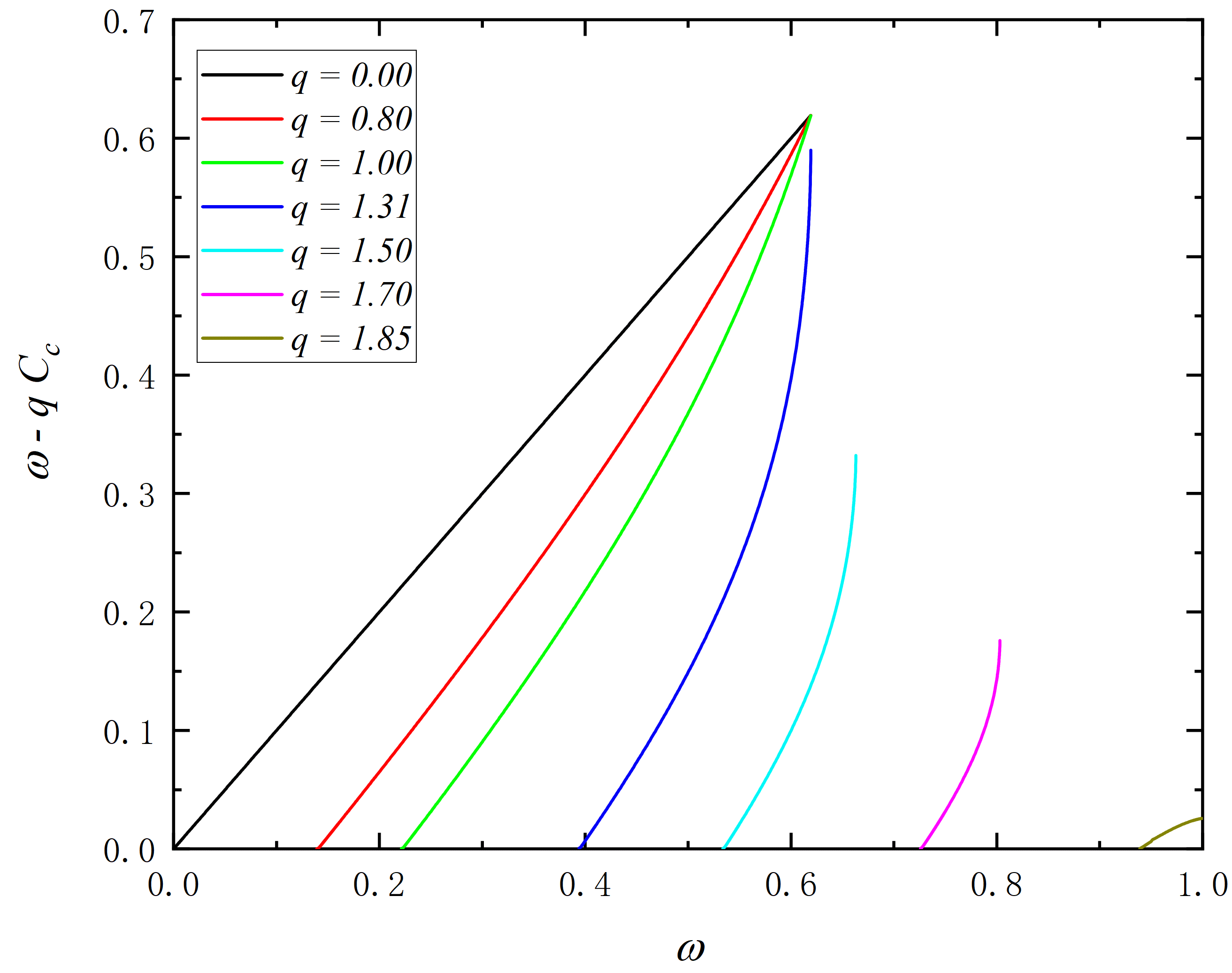}
    \end{center}
    \caption{\small
The ``effective frequency" $ \omega - q c_c $ as a function of the frequency $\omega$. The $p$ of different panels, along with the colors of lines representing $q$, are the same as in figure~\ref{admm}.}
\label{ef}
\end{figure}

In figure~\ref{admm}, a notable issue arises in the right-middle panel (where $ p = 0.70 $), where the cyan curve (for $ q = 1.30 $) is not continuous, with a gap between the first and second branches. To address this, we define an ``effective frequency" $ \omega - q c_c $, $c_c$ is the maximum of the potential function $c$. From the field equations, it is evident that for charged Dirac fields, the combination $ \omega - q c $ effectively replaces $ \omega $ as the frequency. Since the function $ c $ reaches its maximum at the center and has a first derivative of zero, we choose $ \omega - qc_c $ as the effective frequency. In figure~\ref{ef}, we plot the relationship between the effective frequency $ \omega - qc_c $ and the frequency for various values of $ p $ and $ q $, with colors and parameters corresponding to those in figure~\ref{admm}. In all panels, the black lines represent solutions for $ q = 0 $, which appear as a straight line with a slope of $1$, and as $ q $ increases, the curves gradually shift downward and to the right. For smaller values of $ p $, the curves exhibit a spiraling trend. As $ p $ increases, the right-middle panel shows that for smaller values of $ q $, the curves maintain a spiraling trend, while when $ q $ reaches $1.30$, the spiral appears to be submerged by the horizontal line where the effective frequency is zero. Only the regions where the effective frequency is greater than zero have solutions, and as $ q $ continues to increase, only a portion of the first branch exists. As $ p $ increases further, the condition $ \omega - qc_c > 0 $ is satisfied for the first branch across all $ q $. We can conclude that the restriction imposed by the effective frequency causes a discontinuity between the first and second branches for the case where $ p = 0.70 $ and $ q = 1.30 $. An effective frequency greater than zero indicates that the energy density of the Dirac field is positive in our model.

\subsection{Frozen Stars}\label{sSec4}
\begin{figure}[h!]
    \begin{center}
        \includegraphics[height=.22\textheight]{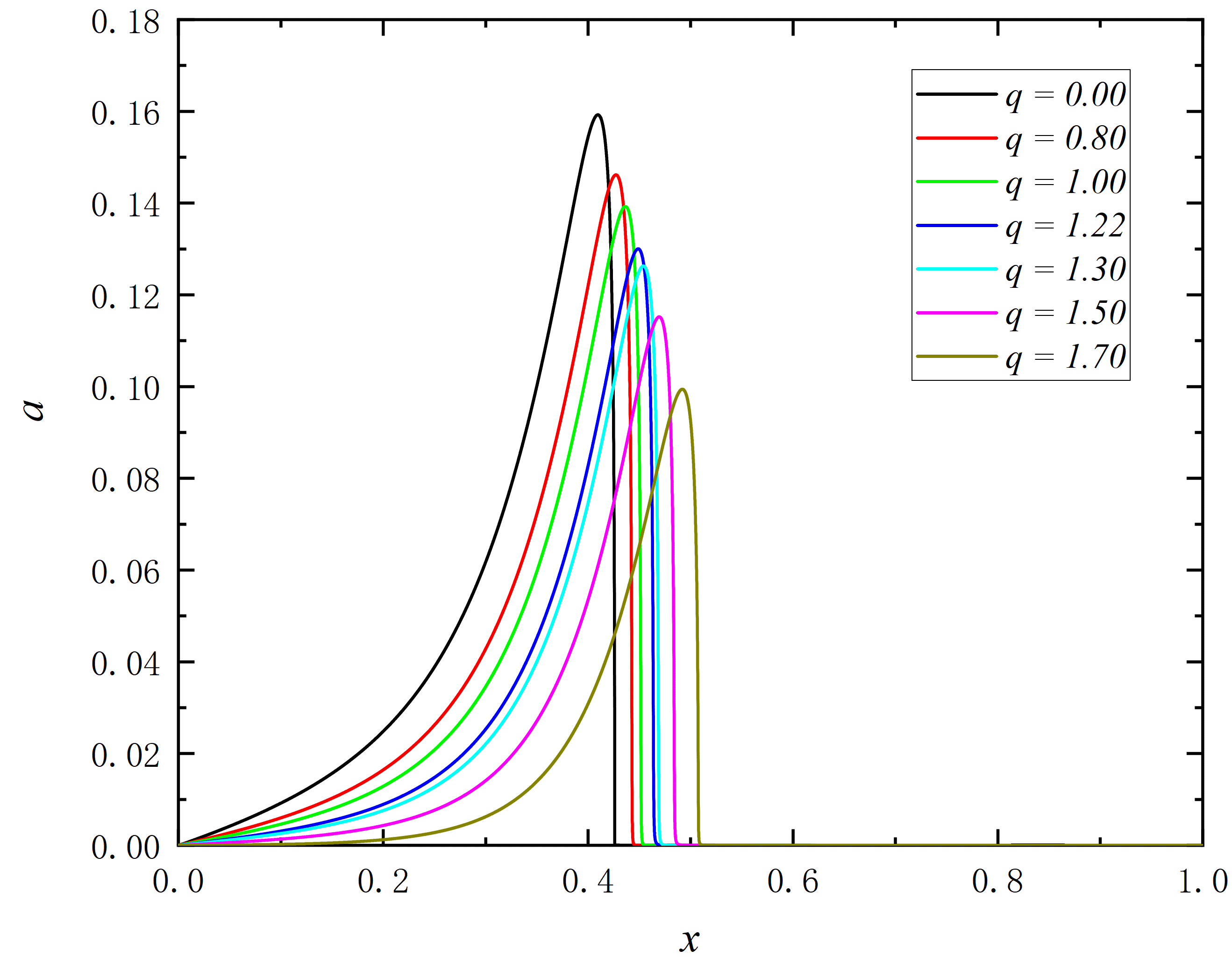}
        \includegraphics[height=.22\textheight]{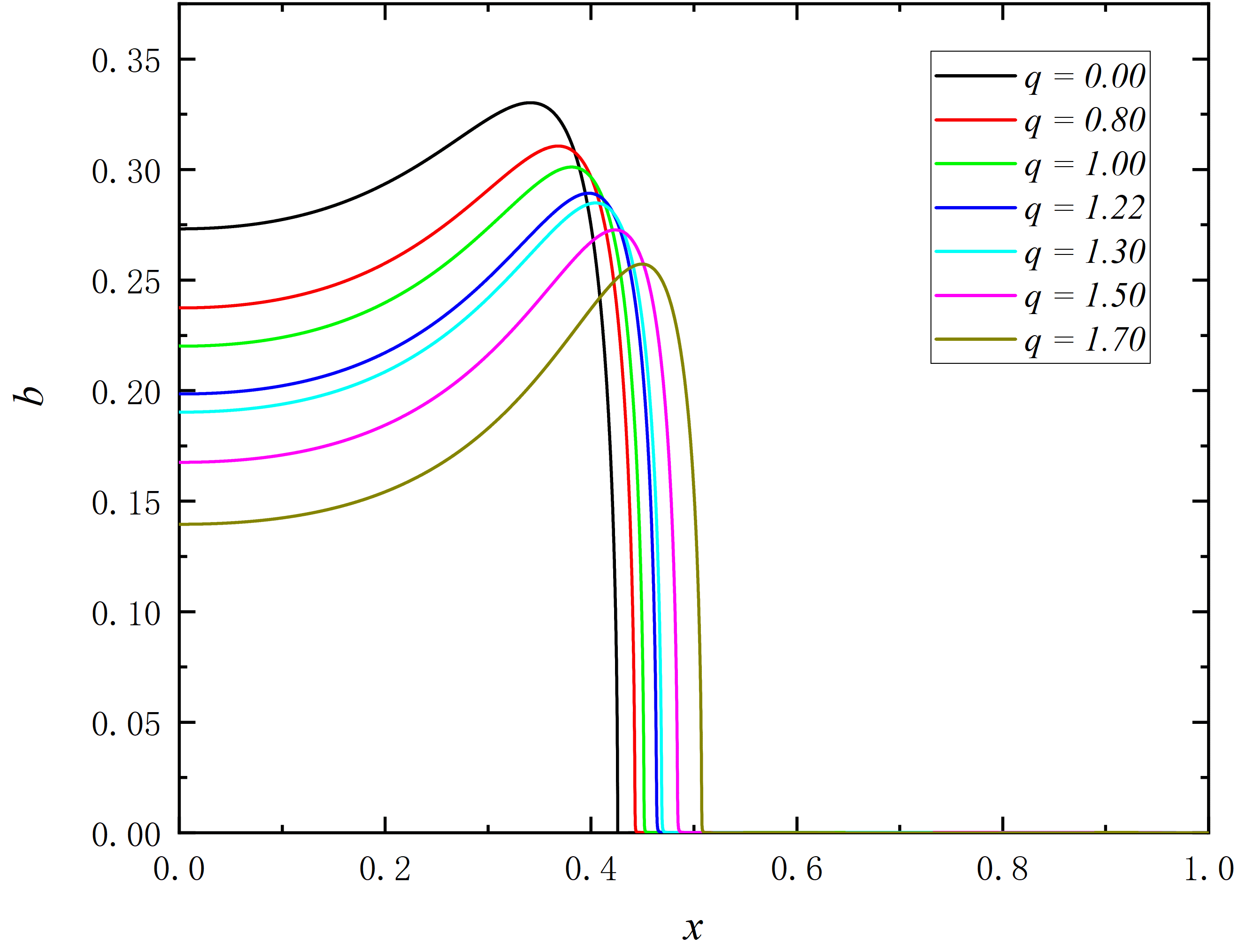}
        \includegraphics[height=.22\textheight]{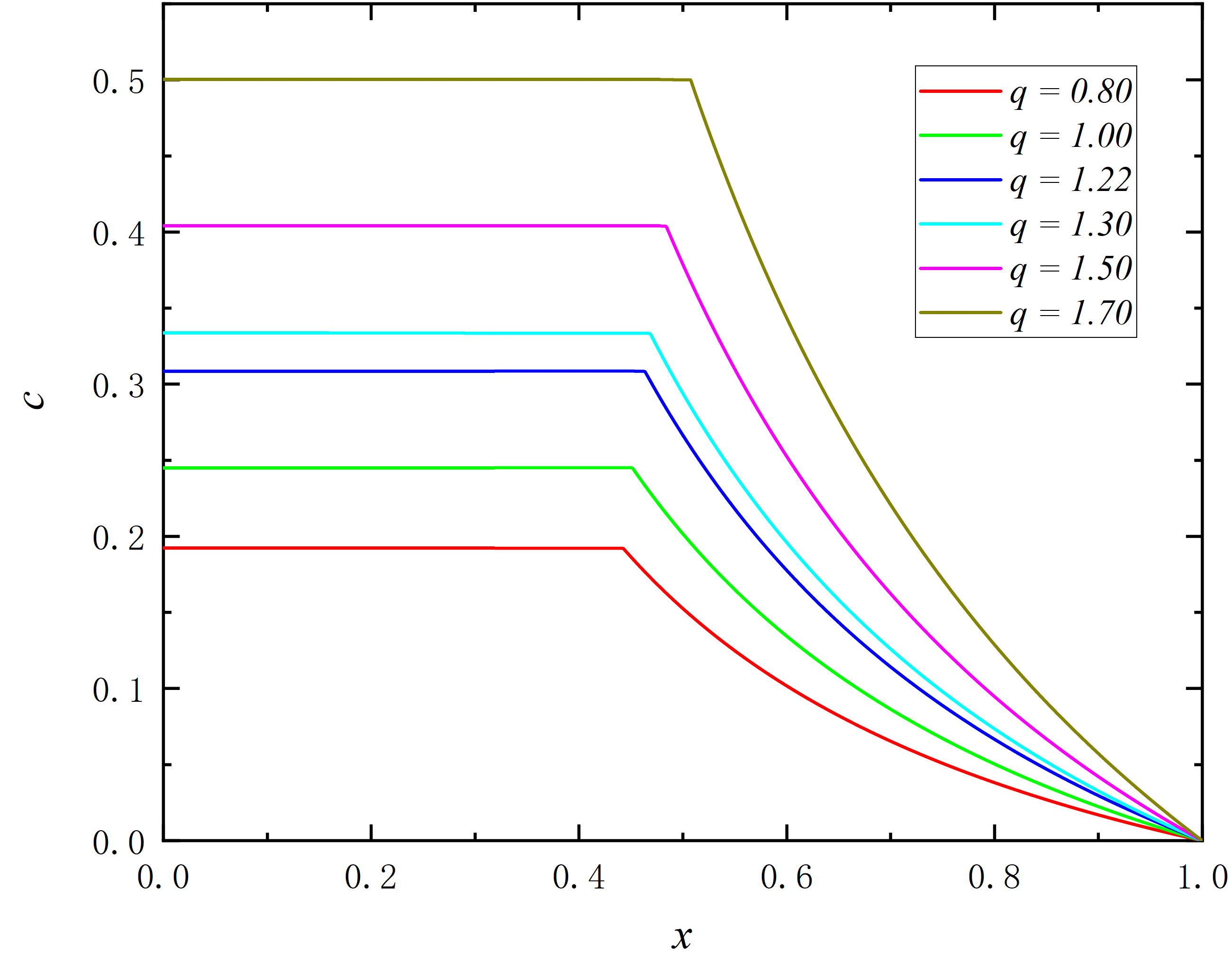}
        \includegraphics[height=.22\textheight]{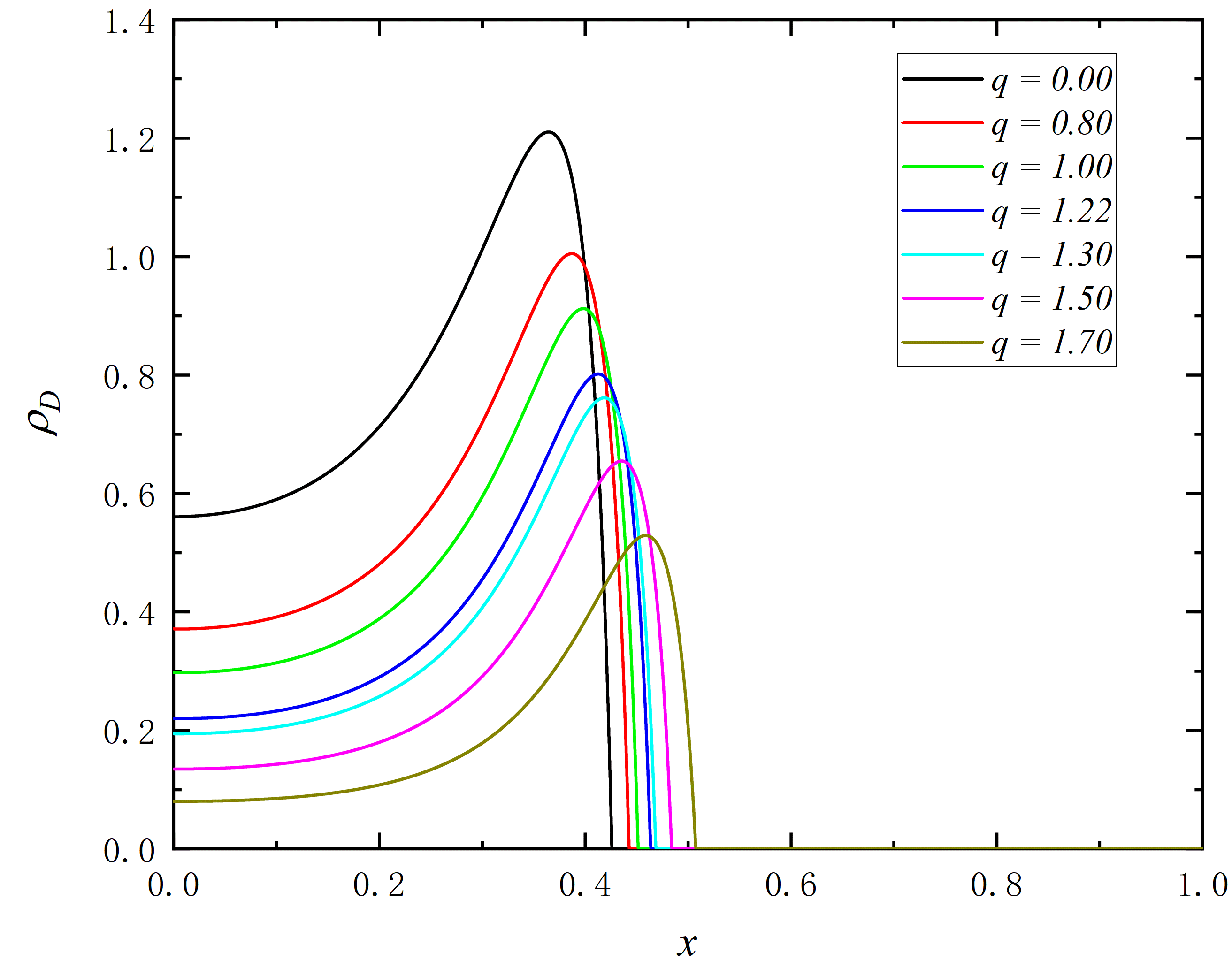}
        \includegraphics[height=.22\textheight]{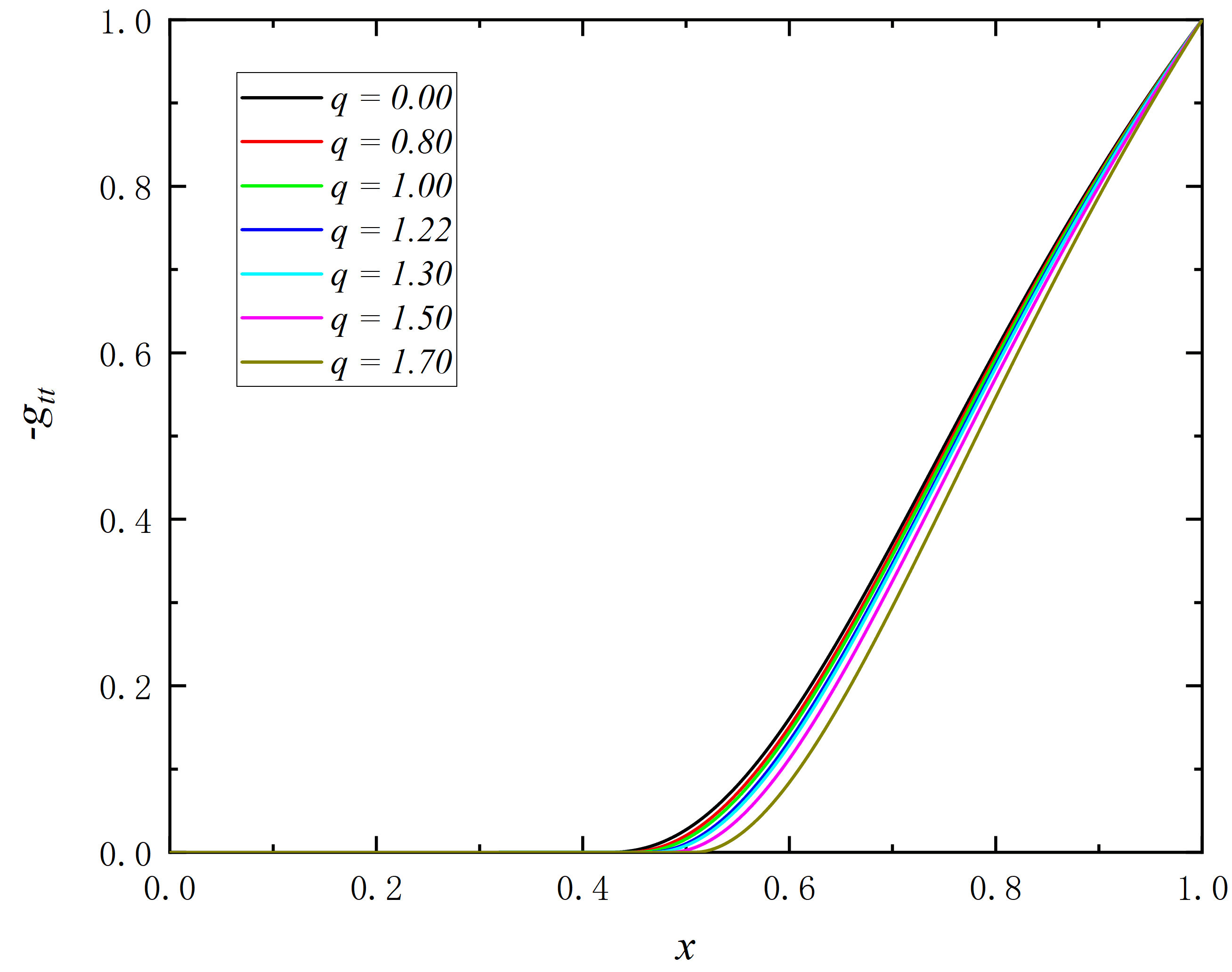}
        \includegraphics[height=.22\textheight]{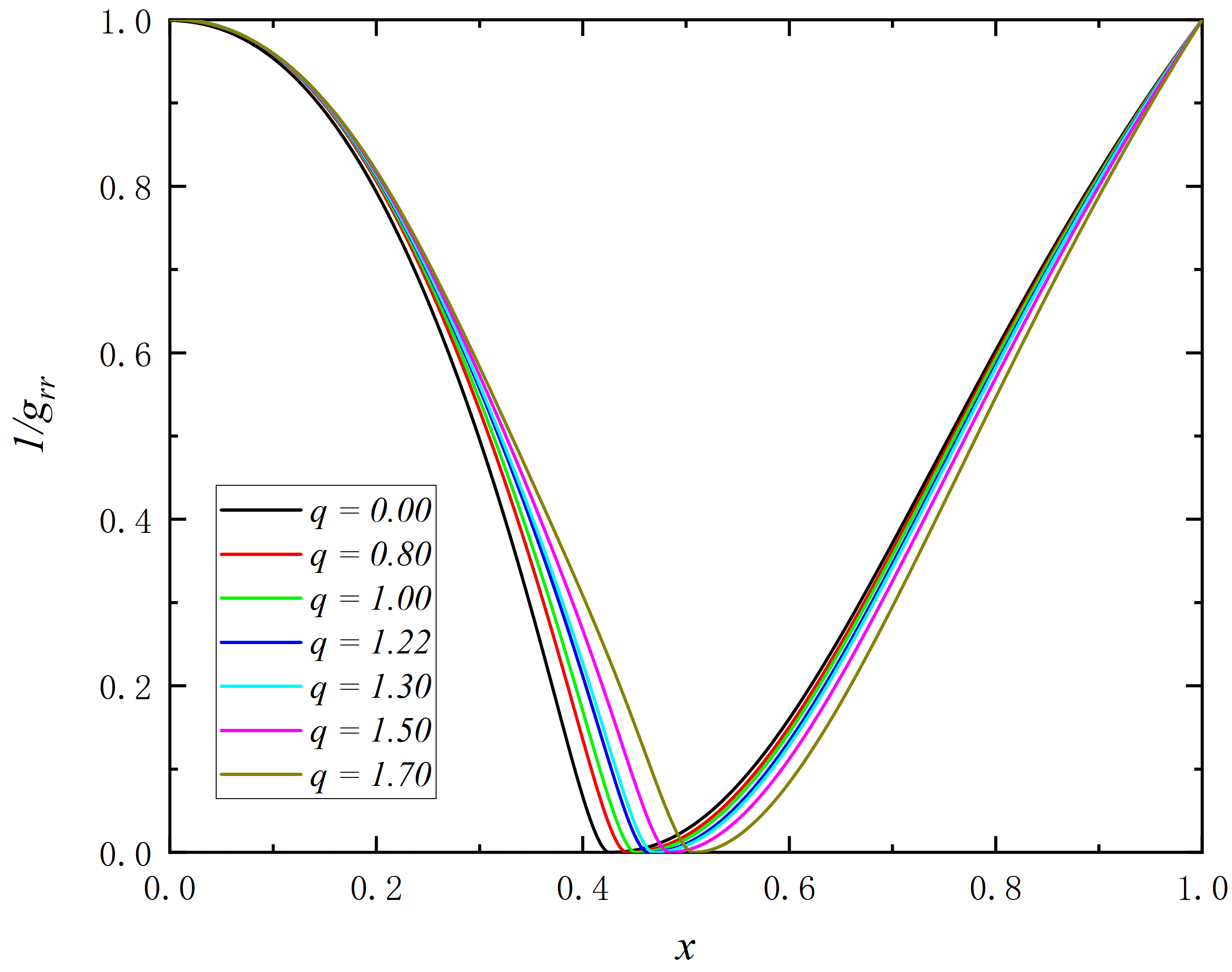}
    \end{center}
    \caption{\small
For $p= 0.75$, the Dirac field functions $a$ and $b$, the electric potential function $c$, the energy density $\rho_D$ of the Dirac field, and the metric components $-g_{tt}$ and $1/g_{rr}$ are distributed in space for different values of $q$.}
\label{fs}
\end{figure}

Reference \cite{Wang:2023tdz} observed that when a scalar or spinor field is introduced into the Bardeen spacetime, the metric component $ g_{tt} $ becomes exceedingly small in a certain region when the field frequency approaches zero, leading to what is termed a ``frozen star". Reference \cite{Wang:2023tdz} found that for charged scalar fields, achieving a frozen star solution does not necessitate the frequency $ \omega $ to be very close to zero, but rather the effective frequency $ \omega - q c_c $ should approach zero. To investigate the effect of charge on the frozen star solution, we present in figure~\ref{fs} the field distributions at $ p = 0.75 $ for different charges, when the effective frequency is very close to zero. It is evident that as the charge $ q $ increases, the peaks of the Dirac field functions $ a $ and $ b $, as well as the energy density of the Dirac fields, progressively diminish and shift outward. For the electrostatic potential function $ c $, the maximum value increases with $ q $. It is observed that for the frozen star solution, at a certain location, $ 1/g_{rr} $ is very small but not zero, and within this location, $ g_{tt} $ is very close to zero. This location is termed the ``critical horizon." The distribution of the Dirac field functions is concentrated within the critical horizon, and the potential $ c $ remains nearly unchanged within the critical horizon, with a sharp inflection point at the critical horizon. Beyond the critical horizon, the electric field strength decays as $ 1/r^2 $, indicating that the charge is concentrated inner the critical horizon. As $ q $ increases, the radius of critical horizon $r_{cH}$ increases.

\subsection{Two Particles Picture}\label{sSec5}
\begin{figure}[h!]
    \begin{center}
        \includegraphics[height=.22\textheight]{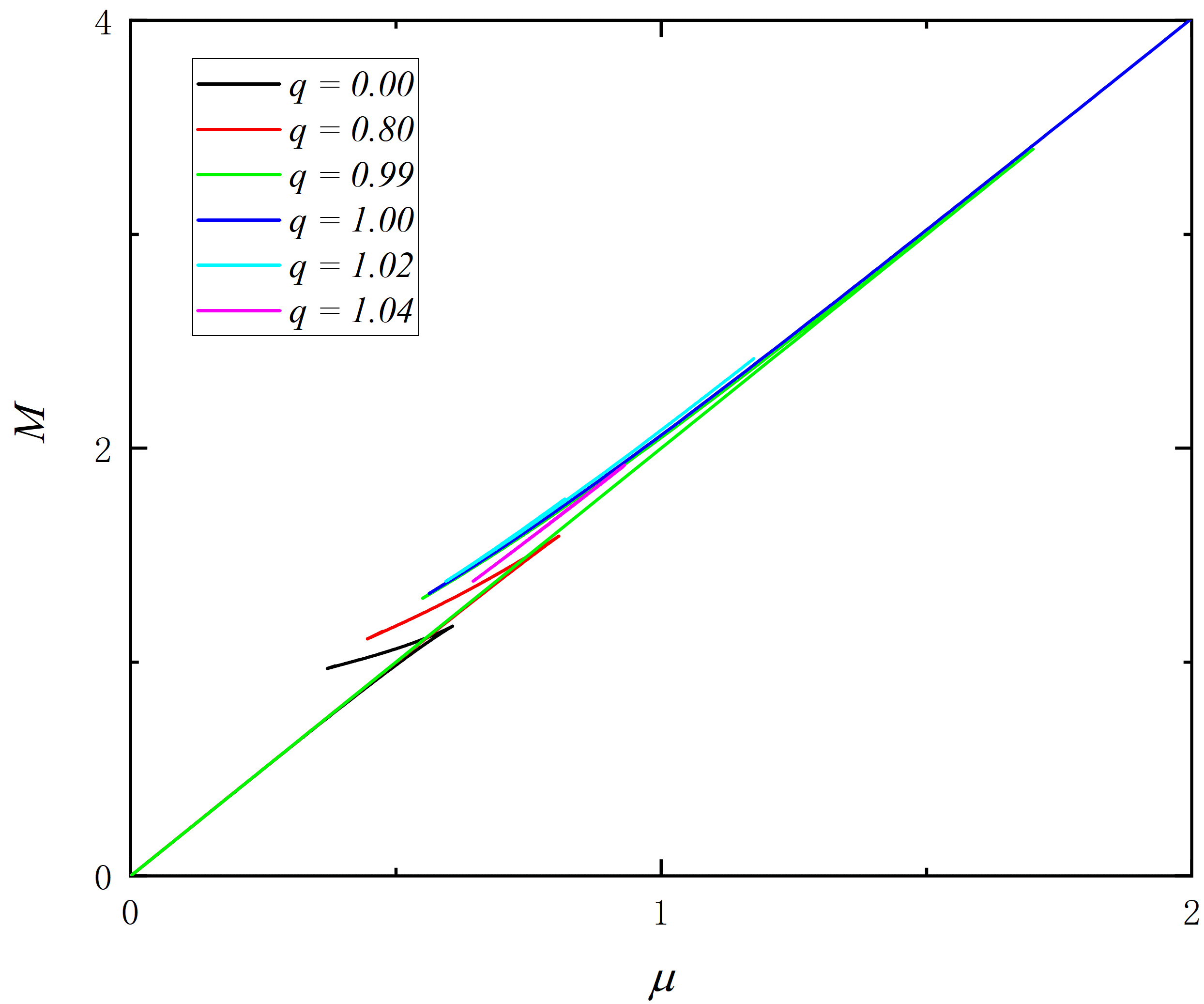}
        \includegraphics[height=.22\textheight]{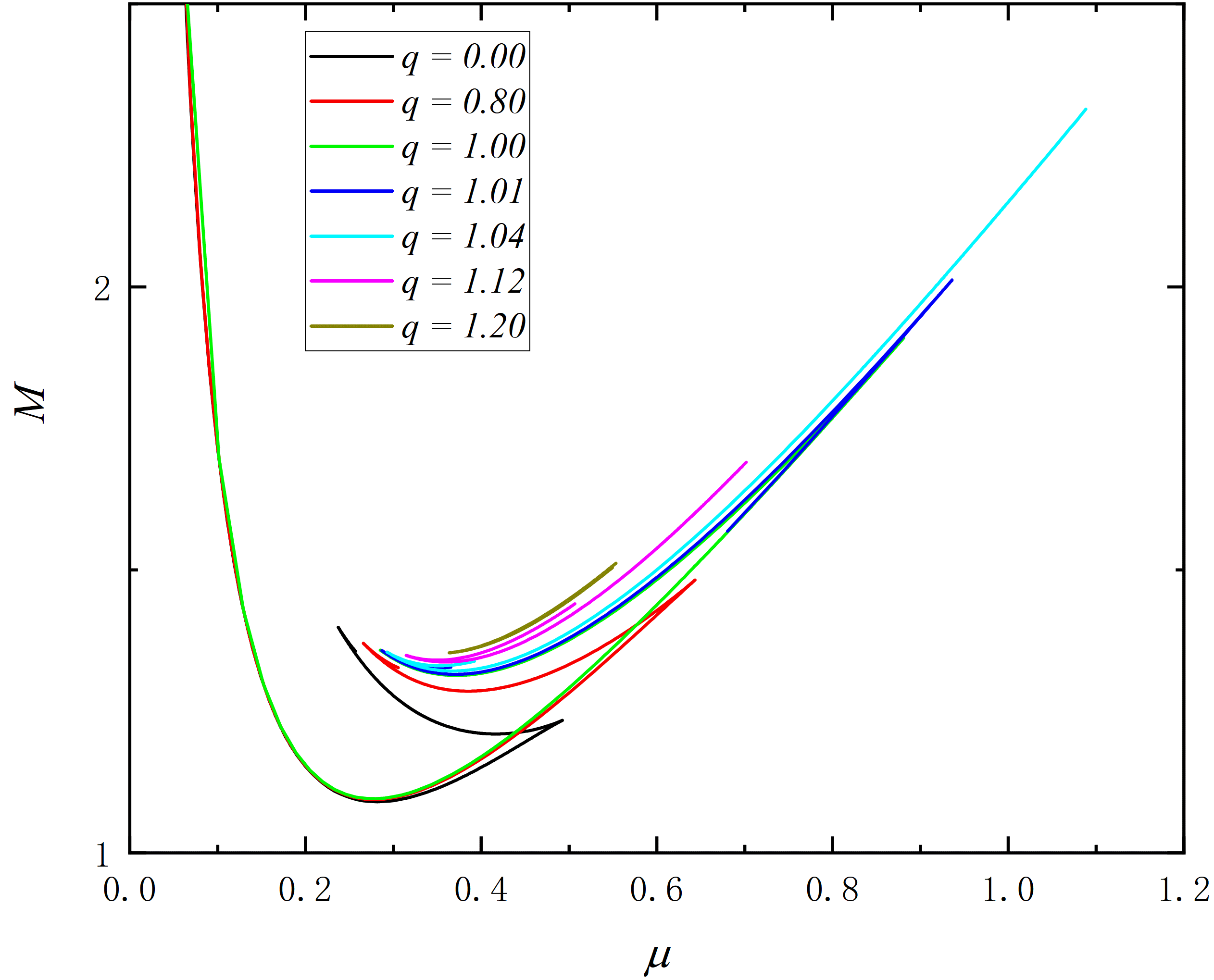}
        \includegraphics[height=.22\textheight]{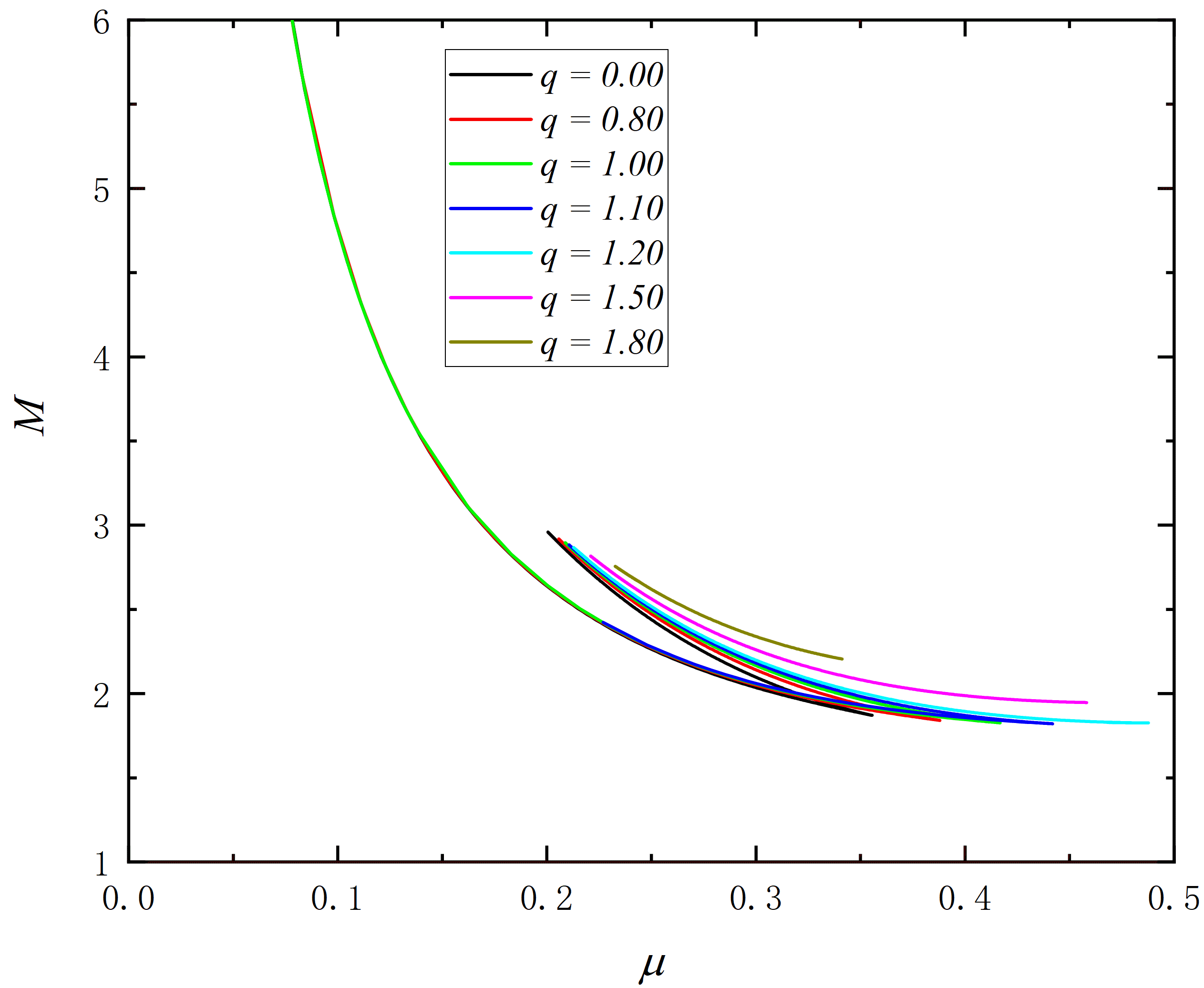}
        \includegraphics[height=.22\textheight]{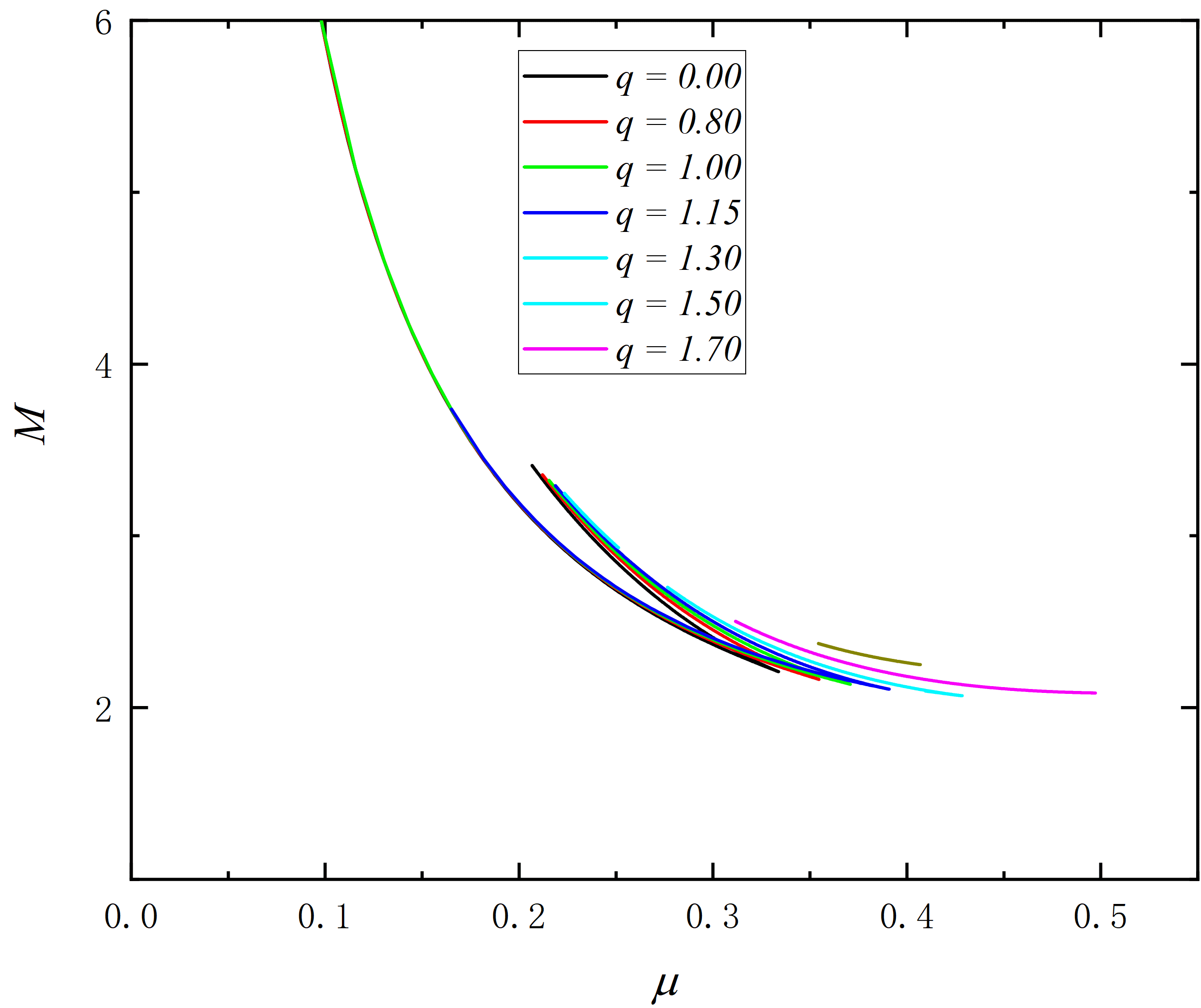}
        \includegraphics[height=.22\textheight]{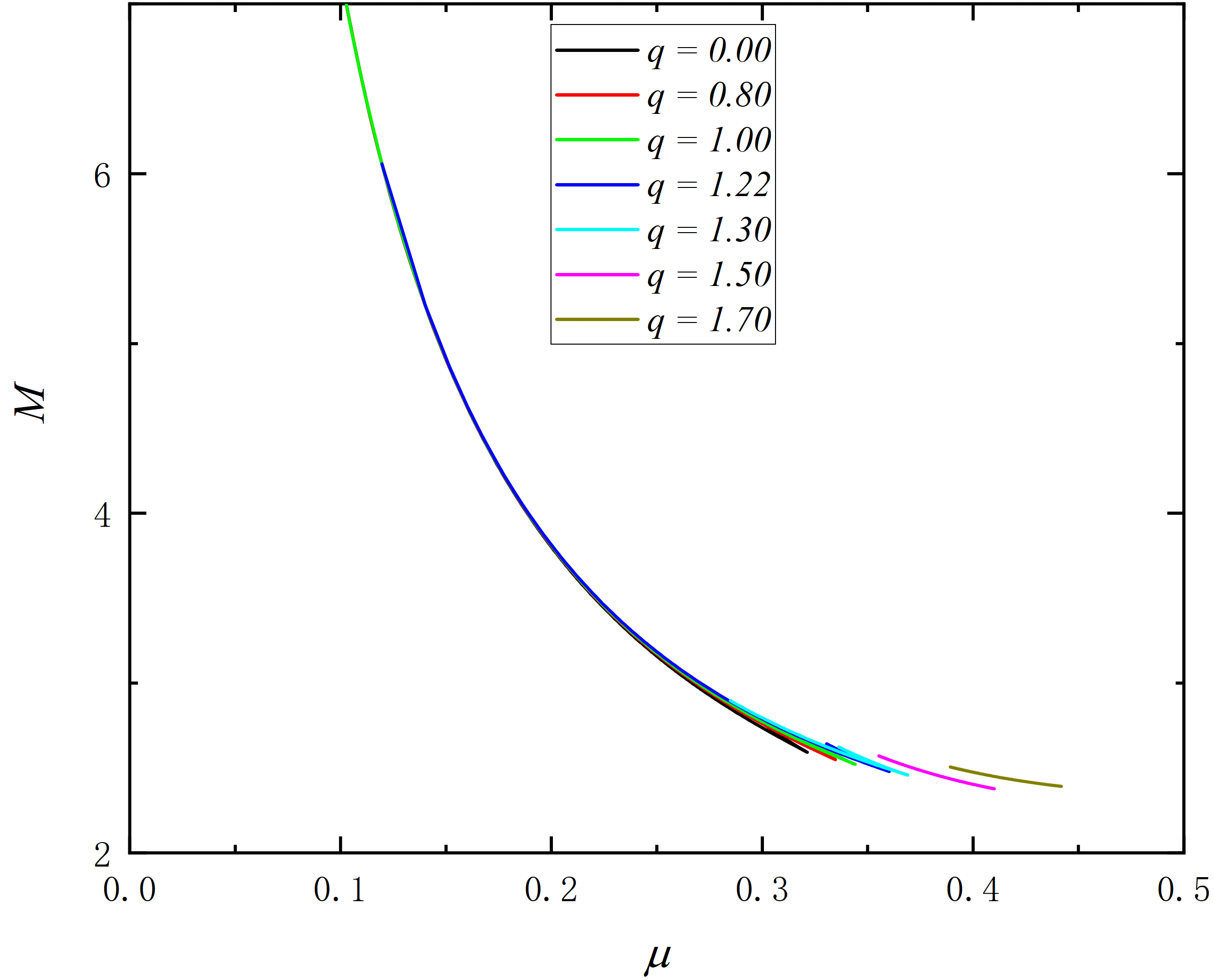}
        \includegraphics[height=.22\textheight]{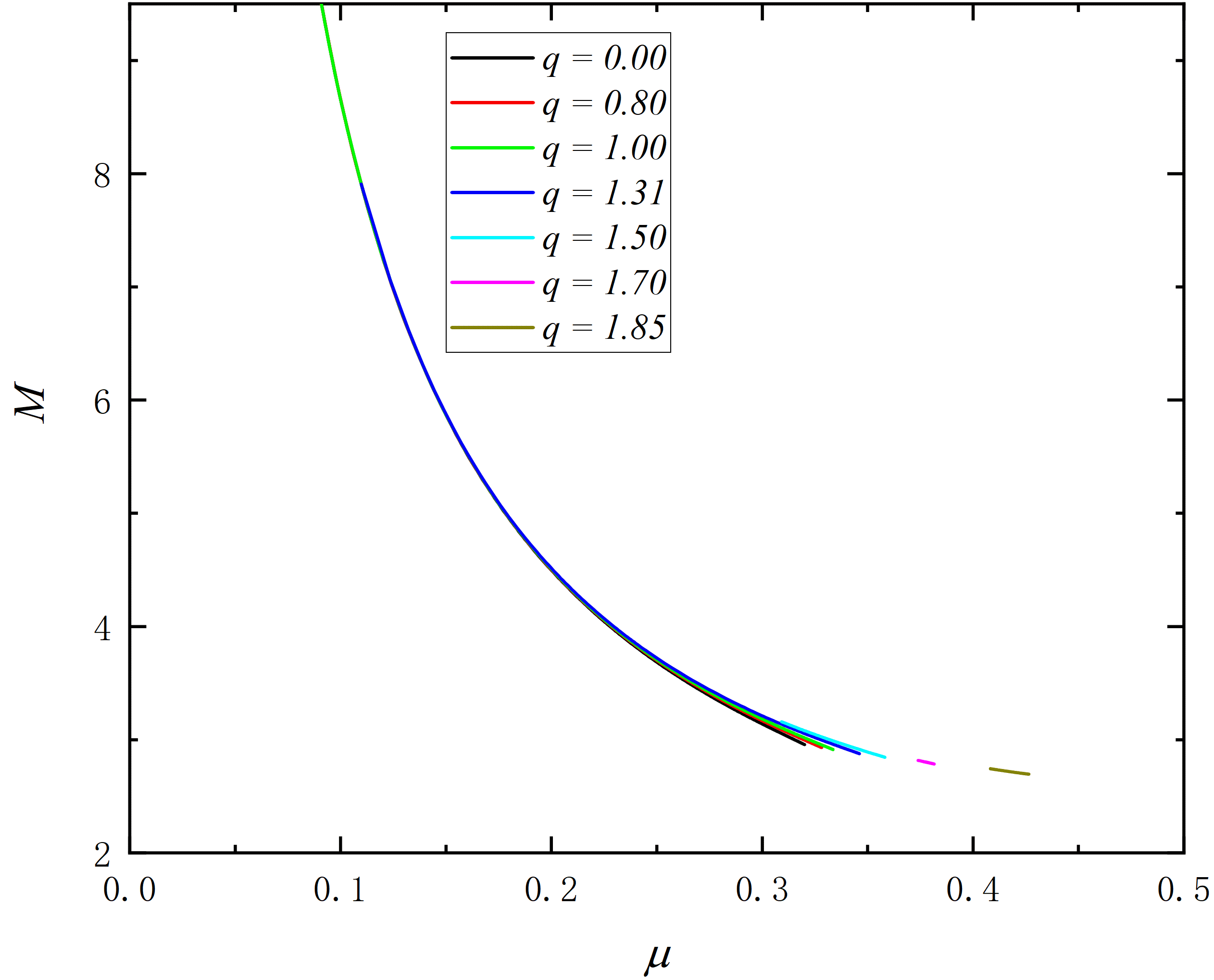}
    \end{center}
    \caption{\small
When the particle number $N=2$ is fixed, the relationship between the ADM mass $M$ and the mass of the Dirac field $\mu$. The $p$ of different panels, along with the colors of lines representing $q$, are the same as in figure~\ref{admm}.}
\label{n2}
\end{figure}

Our results are derived by treating the Dirac field as a classical field, indicating that the particle number is arbitrary. However, the quantum nature of fermions can also be imposed on our results. According to the Pauli exclusion principle, for a system with particle number $ n = 2 $, we establish the relationship between the total mass $ M $ of the spacetime and the fermion mass $ \mu $, which is illustrated in figure~\ref{n2}. It is evident that in the absence of the Bardeen field, as $ \mu $ increases, $ M $ also increases. When $ q $ is less than $ q_c $, the particle mass can be zero, and as $ q $ increases, the maximum achievable particle mass gradually increases. When $ q = q_c = 1 $, the particle mass can be infinitely large, although it cannot be zero. When $ q > q_c $, the particle mass is confined within a certain range, and this range narrows as $ q $ increases. When $p \neq 0$ and $\mu$ are relatively small, the Bardeen field dominates the system's mass, resulting in a total mass much larger than the mass of the Dirac field. For $\mu = 0$, the solution corresponds to a pure Bardeen configuration. As $\mu$ increases, the contribution of the Bardeen field diminishes, leading to a reduction in the system's total mass. When $p$ is small, the system's mass transitions from being dominated by the Bardeen field to being dominated by the Dirac field as $\mu$ increases, causing the total mass $M$ to grow with increasing $\mu$. However, if $p$ is large, the total mass remains dominated by the Bardeen field, regardless of the increase in $\mu$. When $p \neq 0$, the domain of existence of $\mu$ is significantly influenced by the value of $q$.

\section{Conclusion}\label{sec5}

 In this paper, we investigated the model of two charged Dirac fields coupled to Einstein-Bardeen theory in a spherically symmetric spacetime. Through numerical methods, we obtained families of solutions with various magnetic charges and electric charges, analyzing how changes in these two parameters influence the properties of the solutions.

 Firstly, we present the variation of the Dirac field function and the electric potential function with respect to $q$. We observe that, for any fixed frequency $\omega$ and magnetic charge $p$, the Dirac field function $a$ and $b$ exhibit the same variation, while the variation of the electric potential function $c$ is influenced by the magnetic charge $p$. Subsequently, we illustrate how the ADM mass $M$ varies with frequency for different values of $p$ and $q$, noting that all curves retain a spiraling trend. Additionally, we summarize three key factors that significantly affect the shape of the ADM frequency curves: critical charge, maximum frequency, and magnetic charge. If $q$ exceeds the critical charge, the solution cannot revert to pure Bardeen spacetime. If the maximum frequency $\omega_{max} = 1$, the mass of the first branch increases monotonically with frequency; moreover, if $p$ is sufficiently large, the curve lacks a second branch.

 We define an effective frequency and find that for all solutions, the effective frequency is positive. In reference \cite{Wang:2023tdz}, a frozen star solution is obtained when the frequency approaches zero; in our charged model, this translates to the appearance of a frozen star solution when the effective frequency approaches zero. We present the field configurations and metric functions for frozen star solutions with varying magnetic charges, discovering that they exhibit some similarities to uncharged frozen stars. The $g_{rr}$ component approaches zero at a certain location (which we designate as the critical horizon) but does not fall below zero, while the Dirac field is concentrated within the critical horizon. Furthermore, when the magnetic charge is fixed, an increase in $q$ results in a larger critical horizon radius for the frozen star.

 This model allows for several intriguing extensions. In this article, we have only considered the ground state of the Dirac field; exploring excited states with more nodes could unveil distinct characteristics. Furthermore, as we have focused on two Dirac fields, we conducted the two-particle picture, yet we could increase the particle count in the system by considering multiple pairs of Dirac fields, akin to $\kappa$ Dirac stars \cite{Sun:2024nuf}. Lastly, we believe that the spacetime of frozen star solutions merits further investigation, particularly studying the motion of test particles within this spacetime, as well as the search for non-spherically symmetric spacetimes analogous to frozen star solutions.

\section*{ACKNOWLEDGEMENTS}\label{ack}

This work is supported by National Key Research and Development Program of China (Grant No. 2020YFC2201503) and the National Natural Science Foundation of China (Grants No.~12275110 and No.~12047501). 	

\section*{Appendix A}\label{apA}
Within this appendix, we shall elucidate the procedure for obtaining the Dirac equation in the context of curved spacetime.  From the metric in equation (\ref{ds2}), we naturally obtain the vierbein :
\begin{equation}
\label{e}
e_{\mu}^{a}=\left(\begin{array}{cccc}
e^{F_0(r)} \quad & 0 \quad & 0 \quad & 0 \\
0 \quad & e^{F_1(r)} \quad & 0 \quad & 0 \\
0 \quad & 0 \quad & r \quad & 0 \\
0 \quad & 0 \quad & 0 \quad & r\sin\theta 
\end{array}\right),
\end{equation}
which satisfies
\begin{equation}
\label{ee}
g_{\mu\nu}=e_{\mu}^{a}e_{\nu a}, \quad \eta_{ab}=e_{a}^{\mu}e_{b\mu}.
\end{equation}
Here, $\eta_{ab}=\text{diag}(-1,1,1,1)$ is the Minkowski metric. Imposing vierbein compatibility 
\begin{equation}
\label{pte}
\partial_{\mu} e_{\nu}^{a}+\omega_{\mu \  b}^{\  a} e_{\nu}^{b} - \Gamma^{\lambda}_{\ \nu\mu} e_{\lambda}^{a}=0,
\end{equation}
where $\Gamma^{\lambda}_{\ \nu\mu}$ is the affine connection. This equation can lead to the spin connection
\begin{equation}
\label{pte}
\omega_{\mu \  b}^{\  a}=e_{\nu}^{a}e_{b}^{\lambda}\Gamma^{\nu}_{\ \mu\lambda}-e_{b}^{\lambda}\partial_{\mu} e_{\lambda}^{a}.
\end{equation}
And we can get the spin connection matrices
\begin{equation}
\label{Gamma}
\Gamma_{\mu}= -\frac{1}{4}\omega_{\mu ab}\hat{\gamma}^{a}\hat{\gamma}^{b}.
\end{equation}
In the flat spacetime, the gamma matrices $\hat{\gamma}^{a}$ which we choosed are as follows:
\begin{equation}
\label{gam}
\hat{\gamma}^{0}=i\sigma_1\otimes \sigma_0 \quad \hat{\gamma}^{1}=\sigma_2\otimes \sigma_0\quad \hat{\gamma}^{2}=\sigma_3\otimes \sigma_1\quad \hat{\gamma}^{3}=\sigma_3\otimes \sigma_2,
\end{equation}
where the symbol $\otimes$ stands for the direct product and
\begin{equation}
\label{pau}
\sigma^{0}=\left(\begin{array}{cc}
1 \quad & 0 \\
0 \quad & 1  
\end{array}\right) \quad \sigma^{1}=\left(\begin{array}{cc}
0 \quad & 1 \\
1 \quad & 0  
\end{array}\right) \quad \sigma^{2}=\left(\begin{array}{cc}
0 \quad & -i \\
i \quad & 0 
\end{array}\right) \quad \sigma^{3}=\left(\begin{array}{cc}
1 \quad & 0 \\
0 \quad & -1  
\end{array}\right),
\end{equation}
are unit matrix and Pauli matrices. We can get the gamma matrices in curve spacetime:
\begin{equation}
\label{egam}
\gamma^{\mu}=e_{a}^{\mu}\hat{\gamma}^{a}.
\end{equation}
It is readily ascertainable that the gamma matrices $\gamma^{\mu}$ and $\hat{\gamma}^{a}$ satisfy the anti-commutation relations:
\begin{equation}
\label{gamgam}
\{\gamma^{\mu},\gamma^{\nu}\}= 2g^{\mu\nu}I_{4}, \quad \{\hat{\gamma}^{a},\hat{\gamma}^{b}\}= 2\eta^{ab}I_{4},
\end{equation}
where $\{A,B\}=AB+BA$.

\providecommand{\href}[2]{#2}\begingroup\raggedright
\endgroup

\end{CJK*}

\begin{thebibliography}{10}

\bibitem{Wang:2023tdz}
X.~E.~Wang,
``From Bardeen-boson stars to black holes without event horizon,''
[arXiv:2305.19057 [gr-qc]].

\bibitem{EventHorizonTelescope:2019dse}
K.~Akiyama \textit{et al.} [Event Horizon Telescope],
``First M87 Event Horizon Telescope Results. I. The Shadow of the Supermassive Black Hole,''
Astrophys. J. Lett. \textbf{875}, L1 (2019)
[arXiv:1906.11238 [astro-ph.GA]].

\bibitem{EventHorizonTelescope:2019uob}
K.~Akiyama \textit{et al.} [Event Horizon Telescope],
``First M87 Event Horizon Telescope Results. II. Array and Instrumentation,''
Astrophys. J. Lett. \textbf{875}, no.1, L2 (2019)
[arXiv:1906.11239 [astro-ph.IM]].

\bibitem{EventHorizonTelescope:2019jan}
K.~Akiyama \textit{et al.} [Event Horizon Telescope],
``First M87 Event Horizon Telescope Results. III. Data Processing and Calibration,''
Astrophys. J. Lett. \textbf{875}, no.1, L3 (2019)
[arXiv:1906.11240 [astro-ph.GA]].

\bibitem{EventHorizonTelescope:2019ths}
K.~Akiyama \textit{et al.} [Event Horizon Telescope],
``First M87 Event Horizon Telescope Results. IV. Imaging the Central Supermassive Black Hole,''
Astrophys. J. Lett. \textbf{875}, no.1, L4 (2019)
[arXiv:1906.11241 [astro-ph.GA]].

\bibitem{EventHorizonTelescope:2019pgp}
K.~Akiyama \textit{et al.} [Event Horizon Telescope],
``First M87 Event Horizon Telescope Results. V. Physical Origin of the Asymmetric Ring,''
Astrophys. J. Lett. \textbf{875}, no.1, L5 (2019)
[arXiv:1906.11242 [astro-ph.GA]].

\bibitem{EventHorizonTelescope:2019ggy}
K.~Akiyama \textit{et al.} [Event Horizon Telescope],
``First M87 Event Horizon Telescope Results. VI. The Shadow and Mass of the Central Black Hole,''
Astrophys. J. Lett. \textbf{875}, no.1, L6 (2019)
[arXiv:1906.11243 [astro-ph.GA]].

\bibitem{Penrose:1964wq}
R.~Penrose,
``Gravitational collapse and space-time singularities,''
Phys. Rev. Lett. \textbf{14}, 57-59 (1965)

\bibitem{Hawking:1970zqf}
S.~W.~Hawking and R.~Penrose,
``The Singularities of gravitational collapse and cosmology,''
Proc. Roy. Soc. Lond. A \textbf{314}, 529-548 (1970)


\bibitem{Bardeen:1968}
J.~Bardeen,
in Proceedings of GR5,
Tiflis, U.S.S.R. (1968).


\bibitem{Ayon-Beato:1998hmi}
E.~Ayon-Beato and A.~Garcia,
``Regular black hole in general relativity coupled to nonlinear electrodynamics,''
Phys. Rev. Lett. \textbf{80}, 5056-5059 (1998)
[arXiv:gr-qc/9911046 [gr-qc]].

\bibitem{Ayon-Beato:1999kuh}
E.~Ayon-Beato and A.~Garcia,
``New regular black hole solution from nonlinear electrodynamics,''
Phys. Lett. B \textbf{464}, 25 (1999)
[arXiv:hep-th/9911174 [hep-th]].

\bibitem{Ayon-Beato:2000mjt}
E.~Ayon-Beato and A.~Garcia,
``The Bardeen model as a nonlinear magnetic monopole,''
Phys. Lett. B \textbf{493}, 149-152 (2000)
[arXiv:gr-qc/0009077 [gr-qc]].


\bibitem{Bronnikov:2000vy}
K.~A.~Bronnikov,
``Regular magnetic black holes and monopoles from nonlinear electrodynamics,''
Phys. Rev. D \textbf{63}, 044005 (2001)
[arXiv:gr-qc/0006014 [gr-qc]].

\bibitem{Dymnikova:2004zc}
I.~Dymnikova,
``Regular electrically charged structures in nonlinear electrodynamics coupled to general relativity,''
Class. Quant. Grav. \textbf{21}, 4417-4429 (2004)
[arXiv:gr-qc/0407072 [gr-qc]].

\bibitem{Ayon-Beato:2004ywd}
E.~Ayon-Beato and A.~Garcia,
``Four parametric regular black hole solution,''
Gen. Rel. Grav. \textbf{37}, 635 (2005)
[arXiv:hep-th/0403229 [hep-th]].

\bibitem{Berej:2006cc}
W.~Berej, J.~Matyjasek, D.~Tryniecki and M.~Woronowicz,
``Regular black holes in quadratic gravity,''
Gen. Rel. Grav. \textbf{38}, 885-906 (2006)
[arXiv:hep-th/0606185 [hep-th]].

\bibitem{Lemos:2011dq}
J.~P.~S.~Lemos and V.~T.~Zanchin,
``Regular black holes: Electrically charged solutions, Reissner-Nordstr\"om outside a de Sitter core,''
Phys. Rev. D \textbf{83}, 124005 (2011)
[arXiv:1104.4790 [gr-qc]].

\bibitem{Balart:2014jia}
L.~Balart and E.~C.~Vagenas,
``Regular black hole metrics and the weak energy condition,''
Phys. Lett. B \textbf{730}, 14-17 (2014)
[arXiv:1401.2136 [gr-qc]].

\bibitem{Balart:2014cga}
L.~Balart and E.~C.~Vagenas,
``Regular black holes with a nonlinear electrodynamics source,''
Phys. Rev. D \textbf{90}, no.12, 124045 (2014)
[arXiv:1408.0306 [gr-qc]].

\bibitem{Fan:2016rih}
Z.~Y.~Fan,
``Critical phenomena of regular black holes in anti-de Sitter space-time,''
Eur. Phys. J. C \textbf{77}, no.4, 266 (2017)
[arXiv:1609.04489 [hep-th]].

\bibitem{Bronnikov:2017sgg}
K.~A.~Bronnikov,
``Nonlinear electrodynamics, regular black holes and wormholes,''
Int. J. Mod. Phys. D \textbf{27}, no.06, 1841005 (2018)
[arXiv:1711.00087 [gr-qc]].

\bibitem{Junior:2023ixh}
J.~T.~S.~S.~Junior, F.~S.~N.~Lobo and M.~E.~Rodrigues,
``(Regular) Black holes in conformal Killing gravity coupled to nonlinear electrodynamics and scalar fields,''
Class. Quant. Grav. \textbf{41}, no.5, 055012 (2024)
[arXiv:2310.19508 [gr-qc]].

\bibitem{Yue:2023sep}
Y.~Yue and Y.~Q.~Wang,
``Frozen Hayward-boson stars,''
[arXiv:2312.07224 [gr-qc]].

\bibitem{Chen:2024bfj}
J.~R.~Chen and Y.~Q.~Wang,
``Hayward spacetime with axion scalar field,''
[arXiv:2407.17278 [hep-th]].

\bibitem{Huang:2023fnt}
L.~X.~Huang, S.~X.~Sun and Y.~Q.~Wang,
``Frozen Bardeen-Dirac stars and light ball,''
[arXiv:2312.07400 [gr-qc]].


\bibitem{Huang:2024rbg}
L.~X.~Huang, S.~X.~Sun and Y.~Q.~Wang,
``Bardeen spacetime with charged scalar field,''
[arXiv:2407.11355 [gr-qc]].

\bibitem{Zhang:2024ljd}
X.~Y.~Zhang, L.~Zhao and Y.~Q.~Wang,
``Bardeen-Dirac Stars in AdS Spacetime,''
[arXiv:2409.14402 [gr-qc]].

\bibitem{Ma:2024olw}
T.~X.~Ma and Y.~Q.~Wang,
``Frozen boson stars in an infinite tower of higher-derivative gravity,''
[arXiv:2406.08813 [gr-qc]].

\bibitem{Reissner:1916cle}
H.~Reissner,
``\"Uber die Eigengravitation des elektrischen Feldes nach der Einsteinschen Theorie,''
Annalen Phys. \textbf{355}, no.9, 106-120 (1916)

\bibitem{Newman:1965my}
E.~T.~Newman, R.~Couch, K.~Chinnapared, A.~Exton, A.~Prakash and R.~Torrence,
``Metric of a Rotating, Charged Mass,''
J. Math. Phys. \textbf{6}, 918-919 (1965)

\bibitem{Hartle:1972ya}
J.~B.~Hartle and S.~W.~Hawking,
``Solutions of the Einstein-Maxwell equations with many black holes,''
Commun. Math. Phys. \textbf{26}, 87-101 (1972)

\bibitem{Maeda:2008ha}
H.~Maeda, M.~Hassaine and C.~Martinez,
``Lovelock black holes with a nonlinear Maxwell field,''
Phys. Rev. D \textbf{79}, 044012 (2009)
[arXiv:0812.2038 [gr-qc]].

\bibitem{Herdeiro:2018wub}
C.~A.~R.~Herdeiro, E.~Radu, N.~Sanchis-Gual and J.~A.~Font,
``Spontaneous Scalarization of Charged Black Holes,''
Phys. Rev. Lett. \textbf{121}, no.10, 101102 (2018)
[arXiv:1806.05190 [gr-qc]].

\bibitem{Jetzer:1989av}
P.~Jetzer and J.~J.~van der Bij,
``CHARGED BOSON STARS,''
Phys. Lett. B \textbf{227}, 341-346 (1989)

\bibitem{Jetzer:1989us}
P.~Jetzer,
``Stability of Charged Boson Stars,''
Phys. Lett. B \textbf{231}, 433-438 (1989)

\bibitem{Pugliese:2013gsa}
D.~Pugliese, H.~Quevedo, J.~A.~Rueda H. and R.~Ruffini,
``On charged boson stars,''
Phys. Rev. D \textbf{88}, 024053 (2013)
[arXiv:1305.4241 [astro-ph.HE]].

\bibitem{Kumar:2017zms}
S.~Kumar, U.~Kulshreshtha, D.~S.~Kulshreshtha, S.~Kahlen and J.~Kunz,
``Some new results on charged compact boson stars,''
Phys. Lett. B \textbf{772}, 615-620 (2017)
[arXiv:1709.09445 [hep-th]].

\bibitem{Collodel:2019ohy}
L.~G.~Collodel, B.~Kleihaus and J.~Kunz,
``Structure of rotating charged boson stars,''
Phys. Rev. D \textbf{99}, no.10, 104076 (2019)
[arXiv:1901.11522 [gr-qc]].

\bibitem{Lopez:2023phk}
J.~D.~L\'opez and M.~Alcubierre,
``Charged boson stars revisited,''
Gen. Rel. Grav. \textbf{55}, no.6, 72 (2023)
[arXiv:2303.04066 [gr-qc]].

\bibitem{Jaramillo:2023lgk}
V.~Jaramillo, D.~Mart\'\i{}nez-Carbajal, J.~C.~Degollado and D.~N\'u\~nez,
JCAP \textbf{07}, 017 (2023)
doi:10.1088/1475-7516/2023/07/017
[arXiv:2303.13666 [gr-qc]].

\bibitem{Finster:1998ux}
F.~Finster, J.~Smoller and S.~T.~Yau,
``Particle - like solutions of the Einstein-Dirac-Maxwell equations,''
Phys. Lett. A \textbf{259}, 431-436 (1999)
[arXiv:gr-qc/9802012 [gr-qc]].

\bibitem{Herdeiro:2021jgc}
C.~Herdeiro, I.~Perapechka, E.~Radu and Y.~Shnir,
``Spinning gauged boson and Dirac stars: A comparative study,''
Phys. Lett. B \textbf{824}, 136811 (2022)
[arXiv:2111.14475 [gr-qc]].

\bibitem{Finster:1998ws}
F.~Finster, J.~Smoller and S.~T.~Yau,
``Particle - like solutions of the Einstein-Dirac equations,''
Phys. Rev. D \textbf{59}, 104020 (1999)
[arXiv:gr-qc/9801079 [gr-qc]].

\bibitem{Herdeiro:2017fhv}
C.~A.~R.~Herdeiro, A.~M.~Pombo and E.~Radu,
``Asymptotically flat scalar, Dirac and Proca stars: discrete vs. continuous families of solutions,''
Phys. Lett. B \textbf{773}, 654-662 (2017)
[arXiv:1708.05674 [gr-qc]].

\bibitem{Dzhunushaliev:2018jhj}
V.~Dzhunushaliev and V.~Folomeev,
``Dirac stars supported by nonlinear spinor fields,''
Phys. Rev. D \textbf{99}, no.8, 084030 (2019)
[arXiv:1811.07500 [gr-qc]].

\bibitem{Herdeiro:2020jzx}
C.~A.~R.~Herdeiro and E.~Radu,
``Asymptotically flat, spherical, self-interacting scalar, Dirac and Proca stars,''
Symmetry \textbf{12}, no.12, 2032 (2020)
[arXiv:2012.03595 [gr-qc]].

\bibitem{Liang:2023ywv}
C.~Liang, J.~R.~Rena, S.~X.~Sun and Y.~Q.~Wang,
``Multi-state Dirac stars,''
Eur. Phys. J. C \textbf{84}, no.1, 14 (2024)
[arXiv:2306.11437 [hep-th]].

\bibitem{Sun:2024nuf}
S.~X.~Sun, S.~Y.~Cui, L.~X.~Huang, T.~F.~Fang and Y.~Q.~Wang,
``$\kappa $-Dirac stars,''
Eur. Phys. J. C \textbf{84}, no.7, 699 (2024)

\end{thebibliography}
\end{document}